\documentclass[fleqn,usenatbib]{mnras}

\usepackage{newtxtext,newtxmath}

\usepackage[T1]{fontenc}
\usepackage{ae,aecompl}

\usepackage{graphicx}	
\usepackage{amsmath}	

\usepackage{amssymb}	
\usepackage{comment}

\usepackage{xcolor}
\usepackage{ulem}

\title[Evolution of disc cavities and eccentricity]{The evolution of large cavities and disc eccentricity in circumbinary discs}

\author[E. Ragusa et al.]{
Enrico Ragusa,$^{1}$\thanks{E-mail: er198@leicester.ac.uk}
Richard Alexander,$^{1}$
Josh Calcino$^{2}$, Kieran Hirsh$^{3}$, Daniel J. Price$^4$
\\
$^{1}$School of Physics and Astronomy, University of Leicester, Leicester, United Kingdom\\
$^{2}$School of Mathematics and Physics, The University of Queensland, QLD 4072, Australia\\
$^{3}$Univ Lyon, Univ Claude Bernard Lyon 1, ENS de Lyon, CNRS, Centre de Recherche Astrophysique de Lyon UMR5574,\\ F-69230, Saint-Genis-Laval, France\\
$^4$ School of Physics \& Astronomy, Monash University, Clayton, Victoria 3800, Australia}

\date{Accepted XXX. Received YYY; in original form ZZZ}

\pubyear{2020}

\begin{document}
\label{firstpage}
\pagerange{\pageref{firstpage}--\pageref{lastpage}}
\maketitle

\begin{abstract}
We study the mutual evolution of the orbital properties of high mass ratio, circular, co-planar binaries and their surrounding discs, using 3D Smoothed Particle Hydrodynamics simulations.
We investigate the evolution of binary and disc eccentricity, cavity structure and the formation of orbiting azimuthal over-dense features in the disc.
Even with circular initial conditions, all discs with mass ratios $q>0.05$ develop eccentricity. We find that disc eccentricity grows abruptly after a relatively long time-scale ($\sim 400\textrm{--}700$ binary orbits), and is associated with a very small increase in the binary eccentricity. When disc eccentricity grows, the cavity semi-major axis reaches values $a_{\rm cav}\approx 3.5\, a_{\rm bin}$.
We also find that the disc eccentricity correlates linearly with the cavity size.
Viscosity and orbit crossing, appear to be responsible for halting the disc eccentricity growth --  eccentricity at the cavity edge in the range $e_{\rm cav}\sim 0.05\textrm{--} 0.35$. Our analysis shows that the current theoretical framework cannot fully explain the origin of these evolutionary features when the binary is almost circular ($e_{\rm bin}\lesssim 0.01$); we speculate about alternative explanations.
As previously observed, we find that the disc develops an azimuthal over-dense feature in Keplerian motion at the edge of the cavity. A low contrast over-density still co-moves with the flow after 2000 binary orbits; such an over-density can in principle cause significant dust trapping, with important consequences for protoplanetary disc observations.
\end{abstract}




\section{Introduction}

Binaries are common in our Universe, and many phases during the formation and evolution of these binaries involve accretion discs. Their appearance in the electromagnetic spectrum depends on the nature of the objects composing the binary (black holes, stars, planets and moons) and the origin of the gaseous material surrounding them. Among these systems, protostellar/protoplanetary systems (star+star/planet) and black hole (BH) binaries (BH+BH) have recently attracted significant attention in the scientific community.

On the one hand, protostellar/protoplanetary systems are the outcome of the gravitational collapse of molecular cloud cores \citep[for reviews, see ][]{pringle1989, maclow2004}. Even when a binary system is formed, not all the cloud material will land on the forming stars, and the remainder will form a disc around the binary. Furthermore, planet-disc interactions will be the result of planet formation facilitated by the growth of dust grains. Planet-disc systems are just binaries with extreme mass ratios.

Black hole binaries are expected to be found both in the supermassive regime (SMBH binaries) in the gas-rich centres of galaxies powering AGN activity \citep{begelman1980}, and in the stellar regime (SBH binaries, the existence of which has been confirmed by the detection of gravitational waves \citealp{LIGOevent1}),  marking the endpoint of the life of massive stars -- outflows during the life of their stellar progenitors throw gas in to the binary surrounds \citep{demink2017,martin2018}. Stellar BH binaries are also expected to be found in the gas-rich central regions of galaxies \citep{stone2017,bartos2017}.

Despite the differences in physical scales between black hole binaries and protostellar binary systems, the gas dynamics is fundamentally the same, and the interaction between binaries and discs appears to obey the same rules.

Conservation of angular momentum during infall on to the binary forces the material to form a disc.
The binary exerts a tidal torque on the disc \citep{lin1979,goldreich1980}, altering its structure by forming a gap \citep{crida2006,duffell2015b,kanagawa2020} or, if the binary mass ratio is sufficiently high, a cavity \citep{cuadra2009,shi2012,dorazio2013,farris2014,miranda2017}. Vice versa, the disc exerts a back-reaction torque on the binary causing evolution of its orbital properties (migration, eccentricity evolution) and also producing characteristic accretion patterns \citep{artymowicz1996,gunther2002,farris2014,young2015a,ragusa2016,munoz2018,teyssandier2020}.

The disc and the binary primarily exchange angular momentum and energy at resonant locations \citep{goldreich1979,goldreich1980}. A number of theoretical studies have been carried out investigating the effects of individual resonances, in order to determine how they contribute to the evolution of the orbital properties (e.g. \citealp{artymowicz1991,goodman2001,rafikov2002a,goldreich2003}).

Numerical studies have focused on the evolution of binary and disc parameters (e.g. \citealp{kley2006,paardekooper2010,dunhill2013,thun2017,kanagawa2018}), probing the behaviour of the system for large secondary-to-primary mass ratios (e.g. \citealp{cuadra2009,roedig2012,dorazio2013,dunhill2015,shi2015,dorazio2016,munoz2018,munoz2020}), as the theory generally relies on the assumption that the mass ratio of the binary, $q$, is $\ll 1$.

Some issues remain poorly understood, in particular the long term evolution.
The theoretical relationship between the cavity truncation radius and binary properties \citep{artymowicz1994,pichardo2005,pichardo2008,miranda2015} appears to not be fully consistent with the numerical simulations on very long time-scales (\citealp{thun2017,ragusa2018}), where in some cases binaries are observed to carve larger cavities than are predicted theoretically. Recently, resonant theory was found in good agreement with numerical simulations by \citep{hirsh2020}, but it failed to predict the cavity size for the circular, co-planar binary case -- on which we focus in this paper.

A number of numerical simulations starting with circular discs and circular binaries show the growth of both binary and disc eccentricity (e.g. \citealp{papaloizou2001,kley2006,dangelo2006,dunhill2013,dorazio2016,ragusa2018}), even though a seed binary eccentricity $e > 0$ is required in order to excite the eccentric Lindblad resonances which drive eccentricity growth \citep{ogilvie2003,goldreich2003}.
Furthermore, for high mass ratios, a crescent shaped over-dense feature orbiting at the edge of the cavity is likely to form for almost any choice of disc parameters \citep{shi2012,farris2014,ragusa2016,miranda2017,ragusa2017,poblete2019}. The physical mechanism(s) responsible for these features, and their long term evolution, are still poorly understood.

This last issue is of particular interest following the observations performed by the Atacama Large Millimetre Array, and other interferometers. These have imaged a number of protostellar discs with cavities (sometimes referred to as transition discs) and prominent non-axisymmetric features \citep{tuthill2002,andrews2011,isella2013,vandermarel2016,boheler2017,vandermarel2018,casassus2018,pinilla2018,vandermarel2019}, whose origin is still being widely discussed (see Sec. \ref{sec:nonaxfeat} for a thorough discussion).

In this paper, we use a set of 3D Smoothed Particle Hydrodynamics (SPH) simulations to explore the mutual evolution of the binary, which is left free to evolve under the action of the forces exerted by the disc, and disc orbital parameters. We place particular emphasis on the evolution of the disc eccentricity and other disc orbital properties, aiming to explain the physical origin of the crescent shaped azimuthal over-dense features in discs surrounding high mass ratio binaries, and understand the mutual interplay between the binary and the evolution of disc eccentricity and cavity truncation radius.
Long timescale 3D simulations (i.e. $t\gtrsim 1000$ binary orbits) performing a similar analysis are not available in the literature. Three dimensional effects might affect the evolution of the eccentricity, as not allowing the material to access the vertical direction forcing it to move in the x-y 2D plane might spuriously increase the orbital eccentricity.

We prescribe a simple locally isothermal equation of state, and we assume the binary and the disc lie on the same plane.
Other studies have been carried out to discuss the effects of misalignment between the disc and the binary (e.g. \citealp{bitsch2013,aly2015,lubow2015,nealon2018, price2018,hirsh2020}) and alternative prescriptions of the disc thermal structure (e.g. \citealp{baruteau2008,bitsch2013b,benitez2015}).

We allow our simulations to evolve long enough to reach the onset of a quasi-steady evolution. However, we note that most of the results presented in this paper focus on the transition between the initial conditions and the quasi-steady state, as we find that this phase lasts long enough to be relevant for the interpretation of the observations.

The paper is structured as follows: we begin with a broad introduction to resonant binary-disc interaction theory and how this affect the disc and binary evolution (Sec. \ref{sec:resointro} and \ref{sec:mutualevo}); In Sec. \ref{sec:numericalsim} we present our numerical simulations; Sec. \ref{sec:results} presents the results from the simulations; we discuss them in Sec. \ref{sec:discussion}; in Sec. \ref{sec:nonaxfeat} we provide a detailed discussion about the implications of our results in the context of protostellar discs, we draw our conclusions in Sec. \ref{sec:conclusion}.

\subsection{Resonant Binary-Disc Interaction}\label{sec:resointro}

Resonant locations (or resonances) are regions in the disc where the binary and the gas orbital frequency have an integer (or rational) ratio. At these locations the time-varying gravitational potential of the binary
excites density waves \citep{goldreich1980}. Waves carry angular momentum and energy that are transferred to the disc through viscous dissipation or shocks \citep{goodman2001}. Resonances are identified by couples of integers $(m,l)$ and come in two broad types: co-rotation resonances, that are located at
\begin{equation}
    R_{\rm C}=\left(\frac{m}{l}\right)^{2/3}a_{\rm bin}
\end{equation}
and Lindblad resonances, located at
\begin{equation}
    R_{\rm L}=\left(\frac{m\pm 1}{l}\right)^{2/3}a_{\rm bin},
\end{equation}
where $\pm 1$ depends on whether they are outer Lindblad resonances (OLR) or inner Lindblad resonances (ILR), respectively.
The efficiency of angular momentum transfer at a given resonance depends on a number of factors \citep{goldreich2003}, such as the mass ratio of the binary, the eccentricity of the binary, the ``type'' of resonance and (for co-rotation resonances only) the disc vortensity gradient.

When the binary is circular, only resonances $(m,m)$ are effective as the intrinsic efficiency of each resonance scales as $e^{|m-l|}$, where $e$ is the binary eccentricity (and not the exponential function).
For this reason $l=m$ resonances are called ``circular'' resonances. Circular corotation resonances all fall at the co-orbital radius of the binary $R_{\rm C}^{m,m}=a_{\rm bin}$ and for this reason they are also referred to as co-orbital resonances.
When the binary is eccentric a new set of resonances, known as ``eccentric'' resonances, becomes effective.

The ratio between the exchange of angular momentum and energy is fixed by the properties of each resonance.
The overall contribution of the interaction between the binary and the disc at resonant locations determines the evolution of the disc structure and binary orbital parameters.
The torque exerted by the binary on the disc causes the formation of a gap, or if the mass ratio is sufficiently high ($q>0.04$, \citealp{dorazio2016})\footnote{We refer to this threshold value for the transition between gap and cavity as for mass ratios $q>0.04$ no stable orbits around Lagrange points L4 and L5 can be found (tadpole and horseshoe orbits).},
a cavity in the disc, and the onset of disc eccentricity  \citep{lubow1991b,dangelo2006}.
The disc exerts a back reaction torque on the binary causing the binary to migrate (change of semi-major axis) and change the orbital eccentricity.

\subsection{Mutual Evolution of Binary and Disc Orbital Properties}\label{sec:mutualevo}

All resonances lying in the circumbinary disc (i.e. outside the binary orbit) cause inward migration of the binary, while inner resonances (within the binary orbit) promote outward migration. Different resonances in the disc provide different contributions to the binary eccentricity evolution \citep{goldreich2003}: outer circular Lindblad resonances (OCLR, i.e. $R_{\rm L}>a_{\rm bin}$) and non-co-orbital eccentric Lindblad resonances (ELR with $R_{\rm L}\neq a_{\rm bin}$) pump the eccentricity; while circular inner Lindblad resonances (ICLR, i.e. $R_{\rm  L}<a_{\rm bin}$), eccentric corotation resonances (ECR) and co-orbital (i.e. ELR with $R_{\rm L}=a_{\rm bin}$) resonances damp it. Furthermore, the evolution of the disc density structure in the region of co-rotation resonances is expected to cause them to saturate \citep{ogilvie2003}, at which point these resonances become ineffective in their eccentricity damping action, allowing the binary eccentricity to grow. The same ELRs expected to pump the binary eccentricity are expected to pump the disc eccentricity, provided again that some initial disc eccentricity is present \citep{teyssandier2016}.

Due to the absence of ELRs in discs surrounding circular binaries, the evolution of the binary eccentricity in principle should not be possible \citep{goldreich2003}. However, a number of numerical works have shown that it is possible for both the binary and the disc to increase their eccentricities, even in the absence of any initial ``seed'' binary or disc eccentricity \citep{papaloizou2001,dunhill2013}. It is important to note that this result is not surprising at all. The concept itself of circular Keplerian orbit by definition does not imply the presence of a binary object at the centre of the system. Thus, initialising the velocities of fluid elements around a binary using the Keplerian velocity already provides a small seed of orbital eccentricity for the disc.
Finally, we note that fixing the binary orbit throughout the length of the simulation -- as often done in previous works -- breaks the conservation of angular momentum, possibly leading to some spurious growth of the disc eccentricity.

In addition to the resonant interaction, secular interactions also affect the evolution of the disc and binary eccentricity \citep{miranda2017,ragusa2018,teyssandier2019} on long time-scales. Secular interaction is not expected to provide long term growth or damping of the disc eccentricity. Secular effects are instead responsible for periodic oscillations of the eccentricity at fixed semi-major axis (exchange of angular momentum but not of energy).
Secular interactions are also responsible for the precession of the longitude of pericentre of both the binary and the disc. Nevertheless, we note that there are some hints that the individual strengths of different oscillation modes (which depend on the disc-to-secondary mass ratio) appear to have some role in determining the very long time-scale growth of the binary eccentricity ($t\gtrsim 10^5$ binary orbits, \citealp{ragusa2018}).

\section{Numerical simulations}\label{sec:numericalsim}

We performed a set of numerical hydrodynamical simulations using the Smoothed Particle Hydrodynamics code \textsc{phantom} \citep{price2017}.

Our setup consists of two gravitationally bound masses $M_1$ and $M_2$ surrounded by a circumbinary disc (a cavity is already excised when the simulation starts). These masses are modeled as sink particles, where gas particles can be accreted \citep{bate1995}. For numerical reasons we start all our simulations with $M_{\rm tot}=M_1+M_2=1$; we use different binary mass ratios $q=M_2/M_1$ that we will detail in Sec. \ref{sec:spanning}. We initialize our binary on circular orbits with separation $a_{\rm bin}=1$. We use $R_{\rm sink}=0.05$ for both sinks. The sinks are free to migrate due to their mutual gravitational interaction, and their interaction with the circumbinary disc.
This enforces conservation of angular momentum throughout the length of the simulation.

We use SPH artificial viscosity to model the physical processes responsible for the angular momentum transfer through the disc (as prescribed in \citealp{price2017}), that results in an equivalent \citet{shakura1973} viscosity. We discuss the parameters we used for this purpose in Sec. \ref{sec:spanning}.

We allow our simulations evolve for $t=2000 \,t_{\rm orb}$, where $t_{\rm orb}=2{\rm \pi}(GM_{\rm tot}/a^3)^{-1/2}$ is the orbital time of the binary. We note that our choice of disc parameters implies a viscous time $t_{\nu}=1.8\cdot 10^4\textrm{--}10^5\, t_{\rm orb}$ for radii $R=1\textrm{--}7$. Evolving the system for such a long timescale is computationally intractable. However, we will see that that after $2000\, t_{\rm orb}$ all our discs reach a quasi-``steady'' state, meaning that no fast variations of the quantities examined throughout the paper are visible in our results at the end of our simulations (see also the end of Sec. \ref{sec:observ}).
We used $N_{\rm part}=10^6$ SPH particles.

\begin{table}
	\centering
	\caption{Summary of the numerical simulations presented in the paper. The reference name for each simulation contains a number, that refers to the binary mass ratio, and a letter, that refers to the disc properties used. Each simulation has been evolving for $N_{\rm orb}=2000$ binary orbits.
	}
	\label{Tab:sims}
	\begin{tabular}{lcccccc} 
		\hline
		Ref. & $q$ & $p$ & $H_0/R_0$ & $\alpha_{\rm ss}$ &$M_{\rm d}/M_{\rm tot}$  & $R_{\rm in}$\\
		\hline
		1A & 0.01 & 1.5 & 0.05 & $5\times 10^{-3}$ & $5\times 10^{-3}$  & $2.0$\\
		2A & 0.05 & 1.5 & 0.05 & $5\times 10^{-3}$ & $5\times 10^{-3}$  & $2.0$\\
		3A & 0.075 & 1.5 & 0.05 & $5\times 10^{-3}$ & $5\times 10^{-3}$  & $2.0$\\
		4A& 0.1 & 1.5 & 0.05 & $5\times 10^{-3}$ & $5\times 10^{-3}$  & $2.0$\\
	    5A & 0.2 & 1.5 & 0.05 & $5\times 10^{-3}$ & $5\times 10^{-3}$  & $2.0$\\
		6A & 0.5 & 1.5 & 0.05 & $5\times 10^{-3}$ & $5\times 10^{-3}$  & $2.0$\\
		7A & 0.7 & 1.5 & 0.05 & $5\times 10^{-3}$ & $5\times 10^{-3}$  & $2.0$\\
		8A & 1.0 & 1.5 & 0.05 & $5\times 10^{-3}$ & $5\times 10^{-3}$  & $2.0$\\
		\hline
		5C & 0.2 & 1.5 & 0.10 & $5\times 10^{-3}$ & $5\times 10^{-3}$  & $2.0$\\

		5E & 0.2 & 1.5 & 0.05 & $10^{-1}$ & $5\times 10^{-3}$ & $2.0$\\
		5Z & 0.2 & 1.5 & 0.05 & $10^{-2}$ & $5\times 10^{-3}$ & $2.0$\\
		5N & 0.2 & 1.5 & 0.03 & $5 \times 10^{-3}$ & $5\times 10^{-3}$  & $2.0$\\
		5O & 0.2 & 3 & 0.05 & $5\times 10^{-3}$ & $5\times 10^{-3}$  & $2.0$\\
		5P & 0.2 & 0.25 & 0.05 & $5\times 10^{-3}$ & $5\times 10^{-3}$ & $2.0$\\
		5H & 0.2 & 1.5 & 0.05 & $5\times 10^{-3}$ & $ 10^{-2}$  & $2.0$\\
		5A3.0 & 0.2 & 1.5 & 0.05 & $5\times 10^{-3}$ & $5\times 10^{-3}$ & $3.0$\\
		\hline
		6A1.5 & 0.5 & 1.5 & 0.05 & $5\times 10^{-3}$ & $5\times 10^{-3}$ & $1.5$\\
		6A1.7 & 0.5 & 1.5 & 0.05 & $5\times 10^{-3}$ & $5\times 10^{-3}$ & $1.7$\\
		6A1.8 & 0.5 & 1.5 & 0.05 & $5\times 10^{-3}$ & $5\times 10^{-3}$ & $1.8$\\
		6A3.0 & 0.5 & 1.5 & 0.05 & $5\times 10^{-3}$ & $5\times 10^{-3}$ & $3.0$\\
		\hline
	\end{tabular}
\end{table}

\subsection{Reference Case}\label{sec:refcase}
In this section we introduce the disc setup that will be referred to as the ``A'' setup throughout the paper (see all Ref. ``A'' in Table \ref{Tab:sims}). The changes to the parameters used in this setup will be detailed in the next section.

The initial circumbinary disc density profile in our simulations extends from $R_{\rm in}=2$ to $R_{\rm out}=7$. For the inner edge of the disc we follow the rule of thumb that tidally induced cavities around circular binaries have $R_{\rm in}\approx 2a$ \citep{pichardo2008}. We also note that $R_{\rm in}$ lies in between the outermost circular Lindblad resonance $(m,l)=(1,1)$ (OCLR, $2:1$ frequency commensurability, $R_{\rm L}^{1,1}=1.59\,a_{\rm bin}$) and the location of the outermost first order $(m,l)=(2,1)$ ELR ($3:1$ frequency commensurability, $R_{\rm L}^{2,1}=2.08\,a_{\rm bin}$).
We prescribe a tapered power-law density profile of the type
\begin{equation}
    \Sigma(R) =\Sigma_0\left(\frac{R}{R_0}\right)^{-p}\exp\left[-\left(\frac{R}{R_c}\right)^{2-p}\right],
\end{equation}
where we use power-law index $p=1.5$, reference radius $R_0=R_{\rm in}=2$ and tapering radius $R_{\rm c}=5$. We choose $\Sigma_0$ in order to have a disc-to-binary mass ratio $M_{\rm d}/M_{\rm tot}=0.005$.
We use a locally isothermal equation of state $c_{\rm s}=c_{\rm s,0}(R/R_0)^{-q_{c_s}}$ with $q_{c_s}=0.25$. We choose $c_{\rm s,0}$ in order to get a disc aspect-ratio $H_0/R_0=0.05$ at the reference radius $R_0$.

Concerning the disc viscosity, we used an artificial viscosity parameter $\alpha_{AV}=0.2$, $\beta=2$ to prevent particle interpenetration, and allowed artificial viscosity to act also on receding particles (as prescribed in \citealp{price2017}). This  viscous setup results in an equivalent \citet{shakura1973} viscous parameter $\alpha_{\rm ss}=0.005$ \citep{lodato2010,price2017}. We increase the value of $\alpha_{\rm AV}$ to obtain larger values of $\alpha_{\rm ss}$ in other setups.

\subsection{Spanning the Parameter Space}\label{sec:spanning}

In order to study how the system reacts to different physical parameters we ran a large number of different simulations (See Table \ref{Tab:sims} for a list of the simulations). We vary the binary mass ratio $q$ between the the following values $q=\{0.01,0.05,0.075,0.1,0.2,0.5,0.7,1\}$ (different numbers in the ``Ref''. column of Table \ref{Tab:sims} represent different values of $q$).
In addition, we also vary some disc properties to test how these affect the dynamics of the system. In each of them one single parameter is changed with respect to the disc reference case ``A'' (different letters in the Ref. column of Table \ref{Tab:sims}).
In particular, in the case ``5C'' a thicker disc with $H/R=0.1$ is used; in the cases labelled as ``5E'' and ``5Z'' the disc is more viscous than in the ``5A'' cases, using $\alpha_{\rm ss}=10^{-1}$ and $\alpha_{\rm ss}=10^{-2}$, respectively. The cases ``5N, 5O, 5P, 5H, 5A3.0'' use a thinner disc $H/R=0.03$, a steeper initial density profile $p=3$, a shallower density profile $p=0.2$, a larger disc mass and a different inner radius $R_{\rm in}=3$, respectively.
Finally, in order to investigate the dependence on the initial conditions, we performed a set of simulations with $q=0.5$ varying the inner disc radius $R_{\rm in}$.
In particular, simulations ``6A1.5, 6A1.7, 6A1.8, 6A3.0''
have $R_{\rm in}=\{1.5,1.7,1.8,3.0\}\,a_{\rm bin}$.

\section{Results}\label{sec:results}

\begin{figure*}
	\includegraphics[trim={1.7cm 0 4cm 1cm},clip,width=\textwidth]{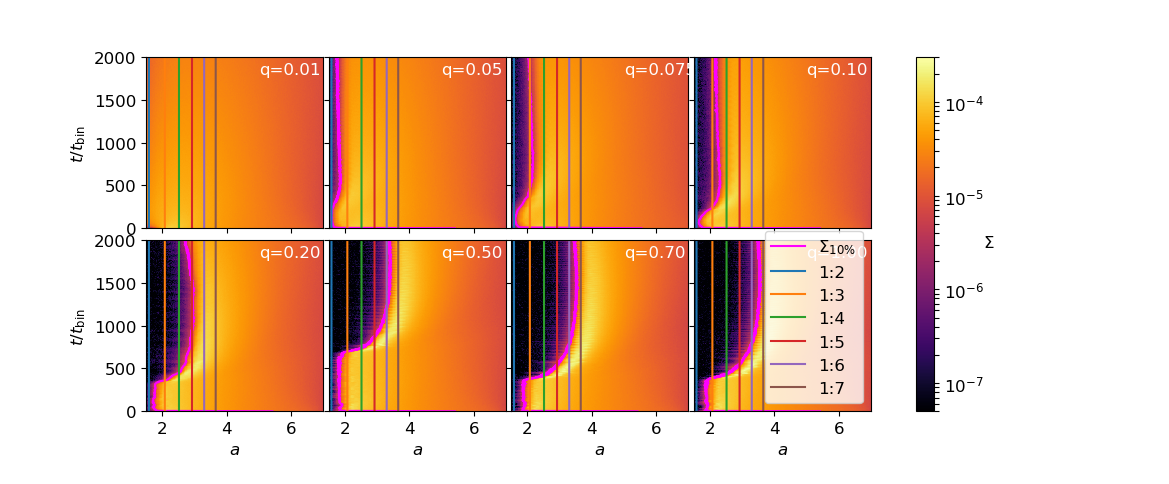}
    \caption{Disc surface density as a function of semi-major axis (x-axis) and time (y-axis) for ``A'' discs with different mass ratios (top left to bottom right $q=\{0.01;0.05;0.1;0.2;0.5;0.7;1\}$, see Tab. \ref{Tab:sims}). The magenta curve superimposed on the plot marks the location of the $10\%$ of the maximum value at each time (i.e. $a_{\rm cav}$ in Eq. (\ref{eq:cavsize})). Vertical lines in different colours mark the location of commensurabilities between the disc and binary orbital frequencies. The main ELRs responsible for eccentricity growth are located at the commensurabilities 1:2 (blue line), $1:3$ (orange line) and $1:4$ (green line). Note the abrupt transition in the cavity structure that takes place after $\approx 400$ binary orbits.}
    \label{fig:simA}
\end{figure*}

\begin{figure*}
	\includegraphics[trim={1.7cm 0 4cm 0.7cm},clip,width=\textwidth]{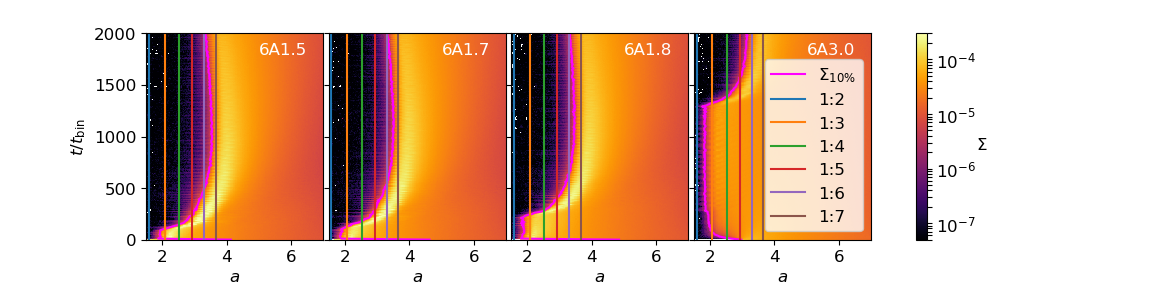}
    \caption{Disc surface density as a function of semi-major axis for 
    a fixed mass ratio $q=0.5$ but different disc initial radii ( simulations  6A1.5, 6A1.7, 6A1.8, 6A3.0  see Tab. \ref{Tab:sims} for the simulations details with $R_{\rm in}=\{1.5,1.7,1.8,3.0\}a_{\rm bin}$, respectively), in order to show that the transition to the eccentric disc configuration occurs earlier when the inner disc is closer to the 1:2 resonance.}
    \label{fig:6SmallLarge}
\end{figure*}

\begin{figure*}
	\includegraphics[width=\textwidth]{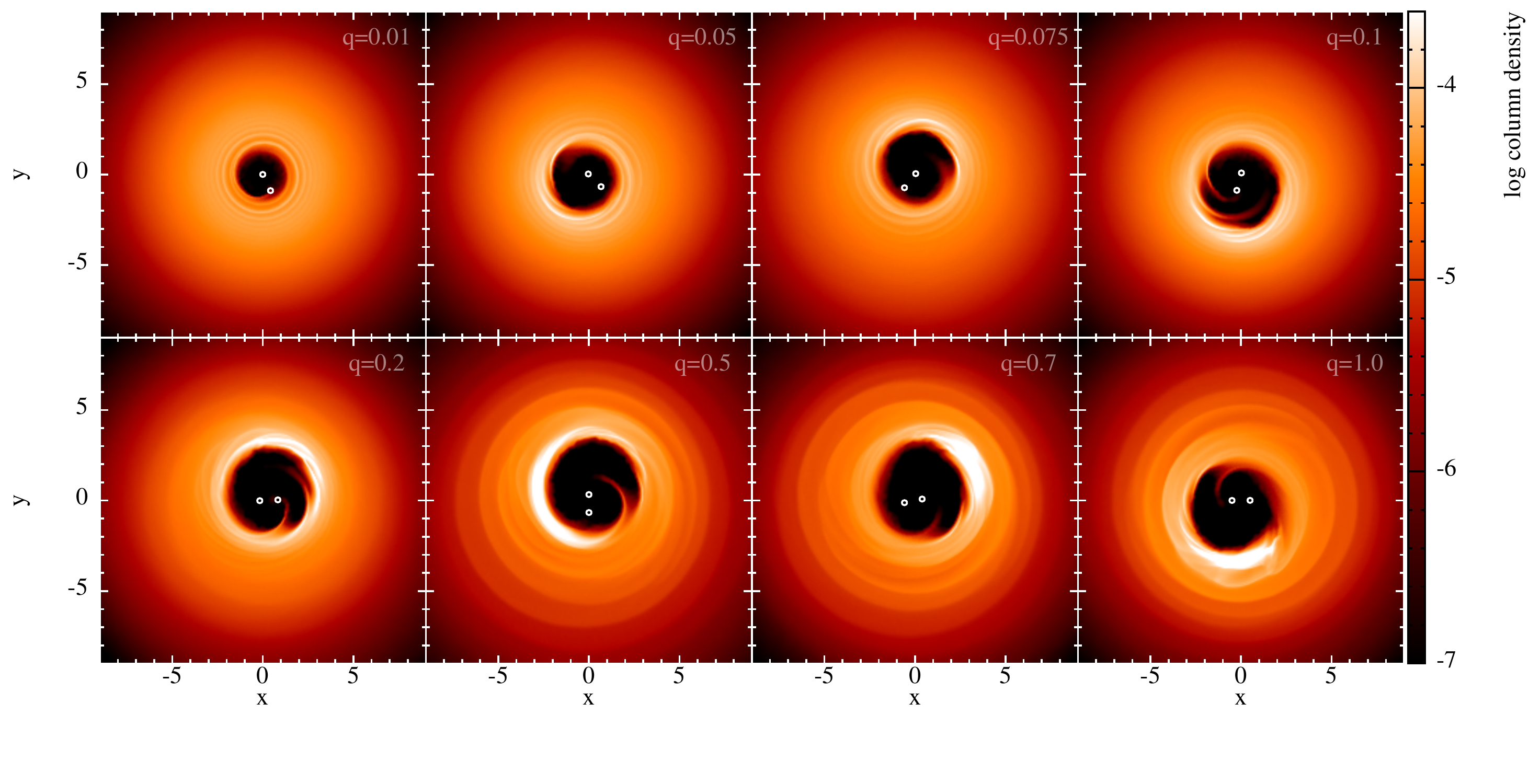}
	\caption{Gas surface density snapshots from simulations ``A'' -- different panels show different binary mass ratios, as detailed in the top right corner of each panel -- after $t\approx 500\,t_{\rm orb}$ (apart from simulation 6A, with $q=0.50$, which is taken at $t=750 \,t_{\rm orb}$, as the transition to an eccentric configuration occurs at later times). The colour scale is logarithmic. An orbiting over-dense lump can be noticed in all simulations with $q\geq 0.2$. }\label{fig:simAs}
\end{figure*}

\begin{figure*}
    \includegraphics[trim={1.7cm 0 4cm 1cm},clip,width=\textwidth]{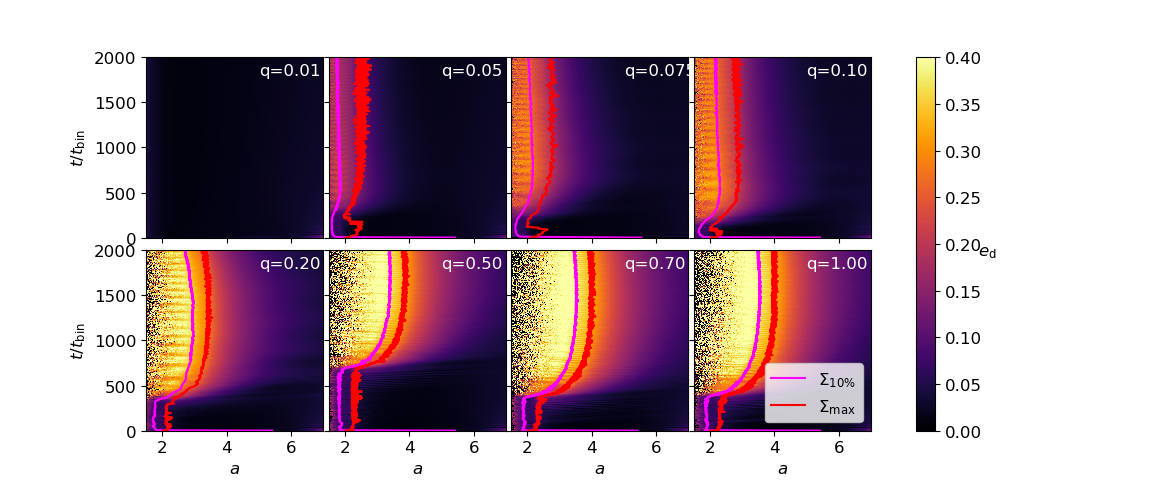}
	\includegraphics[trim={1.7cm 0 4cm 1cm},clip,width=\textwidth]{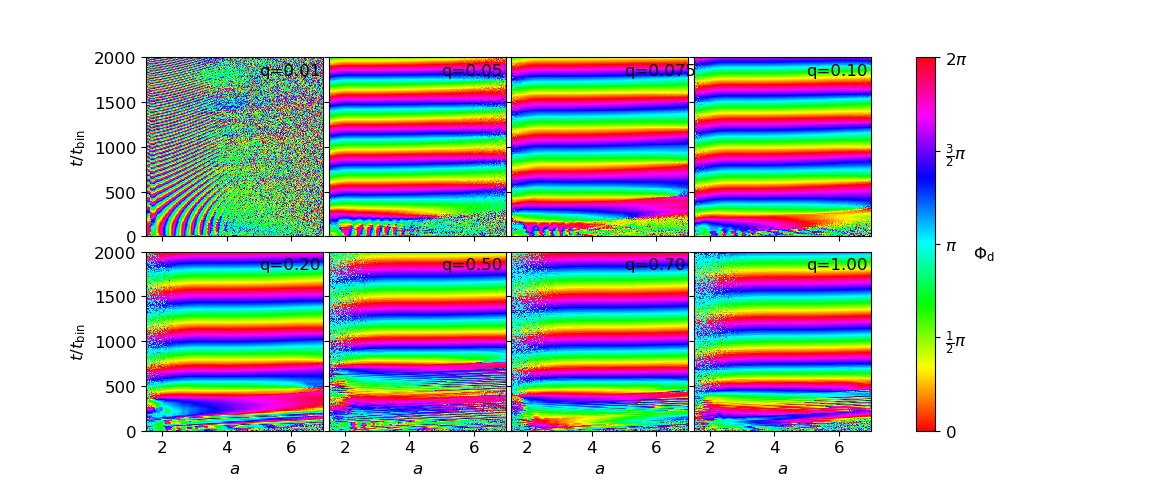}
	\caption{Top panel:  Azimuthally averaged eccentricity profile (colour) as a function of time (y-axis) and semi-major axis (x-axis); the red and magenta curves superimposed to the plot mark the location of the density maximum and location of the $10\%$ of its value at each time (i.e. $a_{\rm cav}$ in Eq. (\ref{eq:cavsize})). Bottom panel: Longitude of the pericentre (colour, azimuthal average) as a function of time (y-axis) and semi-major axis (x-axis) from the simulations: 1A, 2A, 3A, 4A, 5A, 6A and 7A in Table \ref{Tab:sims} ($q=0.2$ see Table \ref{Tab:sims}). We show the main characterising features of the eccentricity evolution: growth of the eccentricity in the disc for $t\gtrsim 400\, t_{\rm orb}$, eccentricity profile decreasing with radius, and disc rigid precession of the pericentre longitude. }\label{fig:pericentre_A}
\end{figure*}

Figure \ref{fig:simA} and \ref{fig:6SmallLarge} summarise how the surface density profile (vertically-integrated volume density) in the disc varies as a function of time in our reference simulations (simulations labelled as ``(1--8)A'') and for different initial inner disc radii (simulations 6A1.5, 6A1.7, 6A1.8, 6A3.0 in Table \ref{Tab:sims}).
These plots show the evolution of the surface density profile $\Sigma(a,t)$ (colours, azimuthal average), as a function of the semi-major axis (x-axis) and time (y-axis). We stress here the importance of producing density profiles using the semi-major axis as a space coordinate instead of radius \citep{teyssandier2017}. When gas orbits in the disc become eccentric, plotting the density as a function of the radius is not ideal, as an element of material spans radii $a(1-e)\leq R\leq a(1+e)$ along its orbit, and this makes it impossible to define the edge of the density profile precisely with a single value of the orbital radius.

\begin{figure*}
	\includegraphics[trim={1.7cm 0.5cm -1.3cm 1cm},clip,width=\textwidth]{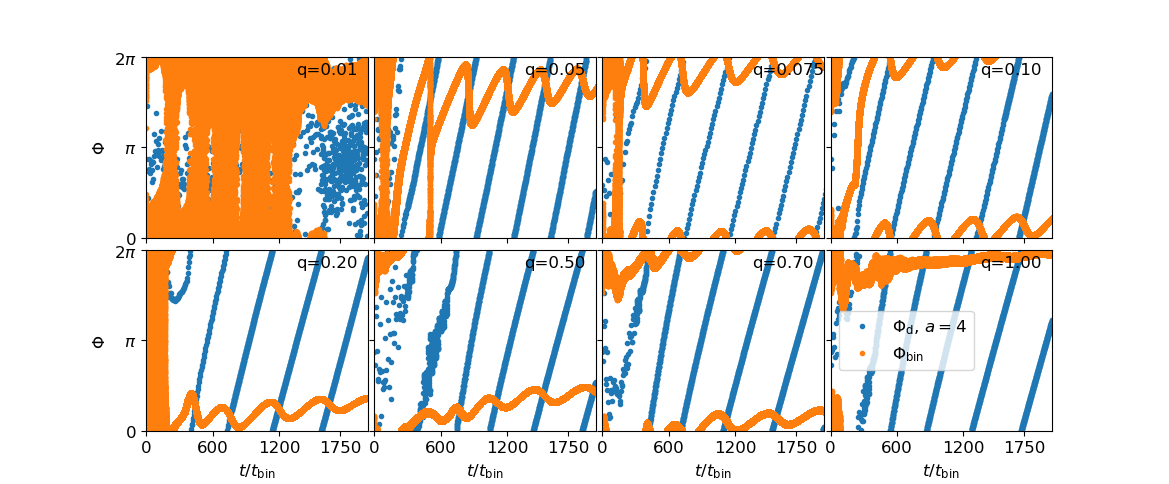}
	\caption{Pericentre phase (y-axis) as a function of time (x-axis) for the disc at $a=4$ (blue dots) and binary (orange dots) for the same simulations in Fig. \ref{fig:pericentre_A}. The disc and the binary start precessing at the same time. The binary pericentre phase precesses at a slower rate than the disc. }\label{fig:pericentre_A2}
\end{figure*}

All our simulations spend  $\sim 400\textrm{--}700\, t_{\rm orb}$ in a ``circular'' steady state, maintaining their circular cavities and without altering their size from the initial configuration. With the exception of the case $q=0.01$, for times $t\gtrsim 400\, t_{\rm orb}$, an abrupt growth of the semi-major axis and eccentricity of the cavity occurs. Furthermore, for mass ratios $q> 0.2$, a prominent azimuthal over-density forms (see Fig. \ref{fig:simAs}), which co-moves with the flow with Keplerian velocity. After this time-scale, the system moves to an ``eccentric'' configuration; the gas orbits consist of a set of nested ellipses with aligned pericentres and an eccentricity profile decreasing with radius (see top panel of Fig. \ref{fig:pericentre_A}). The rigid precession of the disc longitude of pericentre $\Phi_{\rm d}$ -- i.e. the angle the pericentre forms with the positive x axis -- always starts when the transition to the eccentric configuration takes place. We note that this happens because before that time the disc is circular, and it is therefore not possible to attribute any value to the longitude of the pericentre. The binary also starts precessing, although at a much slower rate, as soon as the disc rigid precession starts (see bottom panel of Fig. \ref{fig:pericentre_A2}).

We note here that in a number of previous works the individual masses of the binary are surrounded by circum-primary and circum-secondary discs \citep[e.g.][]{farris2014,ragusa2016,miranda2017} -- usually referred to as ``circum-individual discs'' or ``mini''-discs. Given the relatively low disc viscosity and thickness in our simulations, if these discs form, the low rate at which the binary is fed with the gas from the edge of the cavity makes them progressively sparser, causing SPH numerical viscosity to grow and triggering a positive feedback loop that leads to the disappearance of the circum-individual discs (see Sec. \ref{sec:binsec} for further discussion).

\subsection{Evolution of the Cavity Size}
\begin{figure*}
	\includegraphics[width=0.49\textwidth]{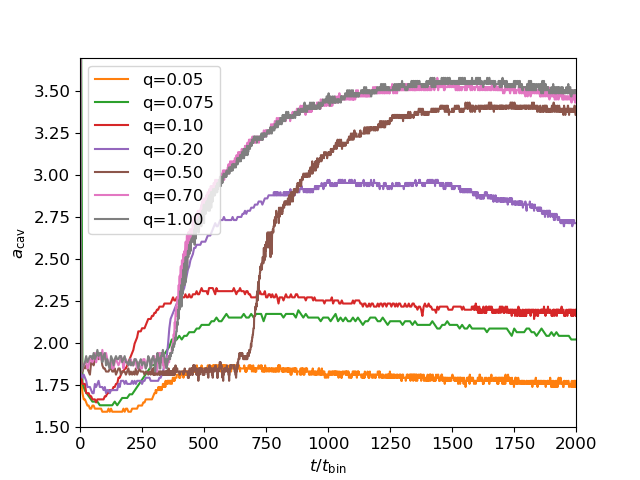}
	\includegraphics[width=0.49\textwidth]{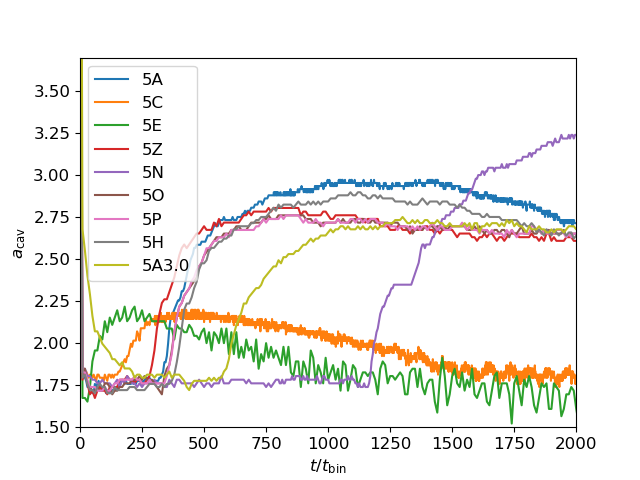}
	\caption{Left panel: Cavity size $a_{\rm cav}$ satisfying Eq. (\ref{eq:cavsize}) as a function of time for the simulations: 1A, 2A, 3A, 4A, 5A, 6A and 7A in Table \ref{Tab:sims}. We deliberately omit the case $q=0.01$ in the left panel of the top row as $a_{\rm cav}$ as the algorithm to solve Eq. (\ref{eq:cavsize}) in order to find $a_{\rm cav}$ of the maximum is not working properly for this case, being very close to the edge of the space domain we use for the analysis of the results ($R=1.5$). Right panel: Same as left panel but for simulations: 5A, 5C, 5E, 5Z, 5N, 5O, 5P, 5H and 5A3.0 (see Table \ref{Tab:sims}). }\label{fig:cav_A}
\end{figure*}

In order to provide a quantitative comparison, we define the cavity size as the semi-major axis at which the value of surface density azimuthal average reaches the $10\%$ of the maximum of the profile at each time, such that
\begin{equation}
    \Sigma(a_{\rm cav},t)\equiv0.1\times \max_a\left[\Sigma(a,t)\right].\label{eq:cavsize}
\end{equation}
We show in the left panel of Fig. \ref{fig:cav_A} the value of $a_{\rm cav}$ as a function of time for our reference simulations (simulations labelled as ``A'' in Tab. \ref{Tab:sims}).

These density profiles were obtained by grouping gas particles in semi-major axis bins, computing the semi-major axis of the $i$-th particle as
\begin{equation}
a_i=-\frac{Gm_iM_{\rm tot}}{2E_i},
\end{equation}
where $E_i$ is the sum of the potential energy and kinetic energy of the $i$-th particle and $m_i$ its mass.
We note that since our estimate of $a_i$ depends on the total mechanical energy of the particle, the velocity corrections due to pressure effects (which we account for when initializing our discs) result in the semi-major axis being slightly underestimated. We note that this discrepancy for our purposes is negligible though, as it scales as $\Delta v_{\rm k}^2\approx (H/R)^2\lesssim 1\%$.

\subsection{Evolution of Disc Eccentricity}
\begin{figure*}
	\includegraphics[width=0.49\textwidth]{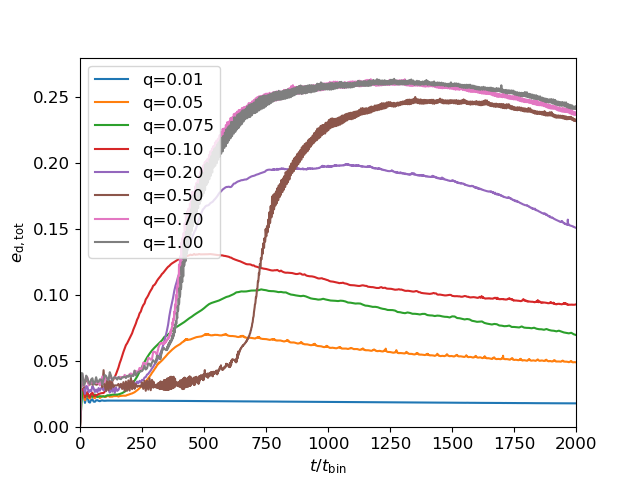}
	\includegraphics[width=0.49\textwidth]{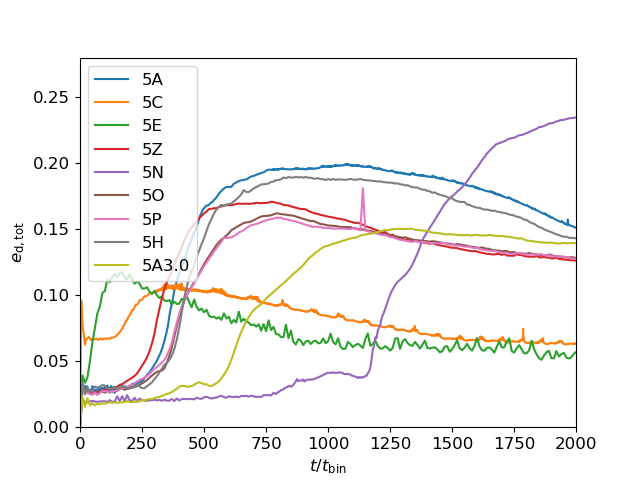}
	\caption{Left panel: Total disc eccentricity $e_{\rm tot}$ in Eq. (\ref{eq:efromAMD2}) as a function of time for the simulations: 1A, 2A, 3A, 4A, 5A, 6A and 7A in Table \ref{Tab:sims}.  Right panel: same as left panel but for the simulations: 5A, 5C, 5E, 5Z, 5N, 5O, 5P, 5H and 5A3.0 (see Table \ref{Tab:sims}). }\label{fig:ecc_A}
\end{figure*}

In order to quantify the disc eccentricity, we define a measure of the ``global'' disc eccentricity as follows.
We compute the total disc angular momentum deficit (AMD) summing the individual contribution of each particle in the disc domain $\mathcal{D}=\{R:1.5\leq R\leq 7\}$ -- a restriction of the disc domain is required as particles with $R\lesssim 1.5$ are no longer moving on Keplerian orbits. -- as follows
\begin{equation}
    {\rm AMD}_{\rm tot}=\sum_{i\in \mathcal{D}}\left(J_{{\rm circ},i}-J_i\right),\label{eq:efromAMD1}
\end{equation}
where the subscript $i$ refers to the $i$-th particle, $J_{{\rm circ},i}=m_{i}\sqrt{a_{i}G M_{\rm tot}}$ is the angular momentum of a particle of mass $m_i$ and semi-major axis $a_{i}$ if it was on a circular orbit, $J_i$ is the particle angular momentum.
We then estimate the ``total'' eccentricity as
\begin{equation}
    e_{\rm tot}=\sqrt{2\frac{{\rm AMD}_{\rm tot}}{\displaystyle \sum_{i\in \mathcal{D}} J_{{\rm circ},i}}}.\label{eq:efromAMD2}
\end{equation}
We plot $e_{\rm tot}$ as a function of time for simulations ``A'' in the left panel of Fig. \ref{fig:ecc_A}.
We remark that this definition provides a global estimate of the disc eccentricity and it is not meant to give a measure of the cavity eccentricity, which is generally higher.

\begin{figure}
	\includegraphics[width=\columnwidth]{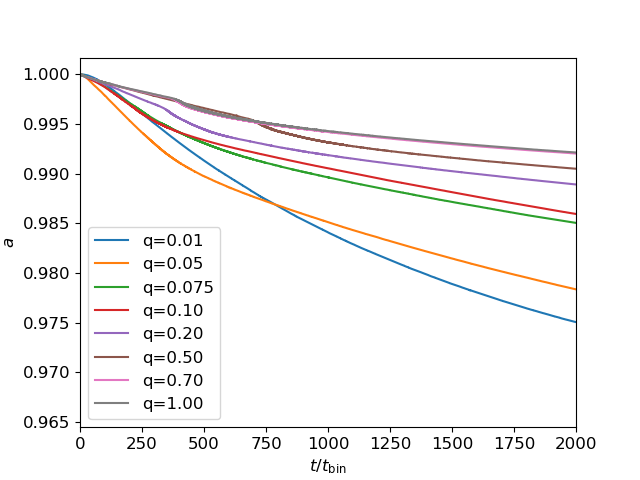}
	\caption{Binary semimajor axis $a_{\rm bin}$ in Eq. (\ref{eq:abineq}) as a function of time for different mass ratios (simulations 1A, 2A, 3A, 4A, 5A, 6A, 7A, 8A in Table \ref{Tab:sims}). We note that the migration rate increases for simulations with $q\geq 0.2$ when the cavity becomes eccentric.}\label{fig:a_As}
\end{figure}

Interestingly, the disc eccentricity grows rapidly until it reaches a maximum value. This value appears to depend on the disc properties and binary mass ratio (a more thorough discussion is provided in Sec. \ref{sec:evodiscecc}). As for the value of $a_{\rm cav}$, we note that, since the total eccentricity $e_{\rm tot}$ is computed using Eq. (\ref{eq:efromAMD1}), the pressure velocity correction results in a small spurious eccentricity $e_{\rm tot,spur}\sim \sqrt{1/2}\times (H/R)$ even when the disc is circular (such that $e_{\rm tot,spur}\sim 3\textrm{--}5\%$ in our simulations).

When in the eccentric configuration, discs always show an eccentricity profile that decreases with radius. As previously mentioned, the disc longitude of pericentre points in the same direction throughout the entire disc, precessing on a time-scale of the order of $100\,t_{\rm orb}$, as shown in Fig. \ref{fig:pericentre_A}.
This implies that the disc behaves rigidly, as originally predicted by \citet{teyssandier2016} and observed later numerically by \citet{miranda2017} and \citet{ragusa2018}.

\subsection{Evolution of the Binary Orbital Parameters}
\label{sec:binsec}

Since our simulations are performed with a ``live'' binary, the back reaction torque the disc exerts on the binary causes the evolution of the orbital properties of the binary.

We compute the binary semi-major axis $a_{\rm bin}$ as
\begin{equation}
a_{\rm bin}=-\frac{GM_1M_2}{2E_{\rm bin}},\label{eq:abineq}
\end{equation}
where $E_{\rm bin}$ is the binary mechanical energy
\begin{equation}
    E_{\rm bin}=\frac{1}{2}M_1v_1^2+\frac{1}{2}M_2v_2^2-\frac{GM_1M_2}{R_{\rm bin}},
\end{equation}
where $R_{\rm bin}=|\mathbf R_2-\mathbf R_1|$ is the physical distance between the two masses and $v_1$ and $v_2$ are velocities computed in the centre of mass (CM) frame. We compute the binary eccentricity $e_{\rm bin}$ as
\begin{equation}
e_{\rm bin}=\sqrt{1-\frac{L_{\rm bin}^2}{\mu^2GM_{\rm tot}a_{\rm bin}}},\label{eq:ebineq}
\end{equation}
where $L_{\rm bin}$ is the total binary angular momentum, in the CM frame, and $\mu=M_1M_2M_{\rm tot}^{-1}$ is the binary reduced mass.

Fig. \ref{fig:a_As} shows the evolution of the binary semi-major axis as a function of time. We note that for $q\geq 0.2$ the evolution of the semi-major axis is characterised by a temporary increase in the migration rate of the binary. We will discuss this effect later in Sec. \ref{sec:horseform}.
The lack of the circum-individual discs surrounding the binary (see the end of Sec. \ref{sec:results}) might impact the evolution of the binary -- some recent works showed that they might produce a positive torque on the binary that lead to outward migration of the binary \citep{tang2017b,munoz2018,moody2019,duffell2019,munoz2020}. However, \citet{tiede2020} and \citet{heath2020} recently showed that migration still occurs inward when the disc aspect-ratio is sufficiently small. More generally, we note that conclusions regarding the evolution of the binary, such as exact migration or eccentricity growth rate, are beyond the scope of this paper. The presence of a live binary mainly allows us to capture secular oscillations of the binary eccentricity, which might play a role in the evolution of the system, and informs us about the intensity of the binary-disc interaction -- when the binary increases its migration rate.

\begin{figure*}
	\includegraphics[width=0.49\textwidth]{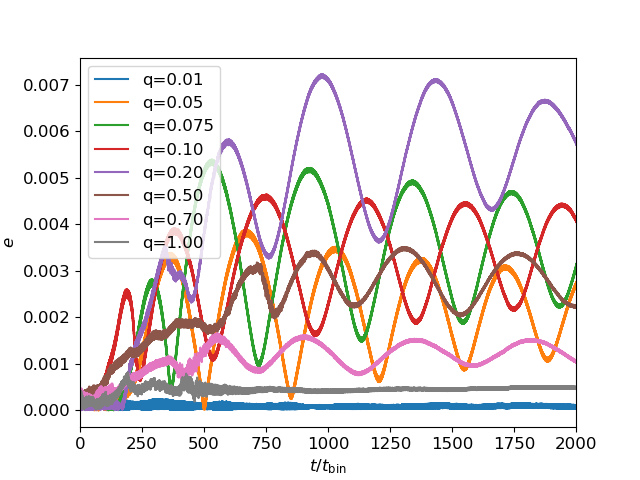}
	\includegraphics[width=0.49\textwidth]{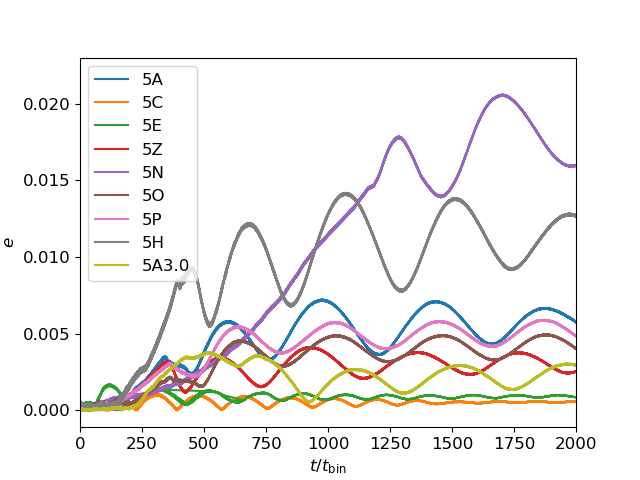}
\caption{Left panel: binary eccentricity $e_{\rm bin}$ in Eq. (\ref{eq:ebineq}) as a function of time for simulations 1A--8A (left panel) and for simulations with mass ratio $q=0.2$ and different disc parameters (right panel). }\label{fig:ebin}
\end{figure*}

The left panel in Fig. \ref{fig:ebin} shows the evolution of the binary eccentricity as a function of time for different binary mass ratios of our reference disc simulations. Secular oscillations of the binary eccentricity can be seen for all cases made exception for $q=0.01$ and $q=1$. The first is consistent with the fact that both disc and binary remain circular for $q=0.01$, and no oscillations can take place. The second, with the fact that equal mass binaries ($q=1$) are not expected to show secular oscillations of the binary eccentricity \citep{miranda2017}.

Fig. \ref{fig:pericentre_A2} shows the evolution of the longitude of pericentre of the binary (orange lines) compared to the evolution of the disc one (blue line, we remark that the disc precesses rigidly, i.e. same pericentre longitude at all radii). It is interesting to note that the precession rate is much lower in the binary than in the disc. The existence of two eigenfrequencies for the precession rate
has been discussed and interpreted through a simplified toy-model in \citet{ragusa2018}.

\subsection{Evolution of azimuthal over-dense features}

\begin{figure*}
	\includegraphics[width=\columnwidth]{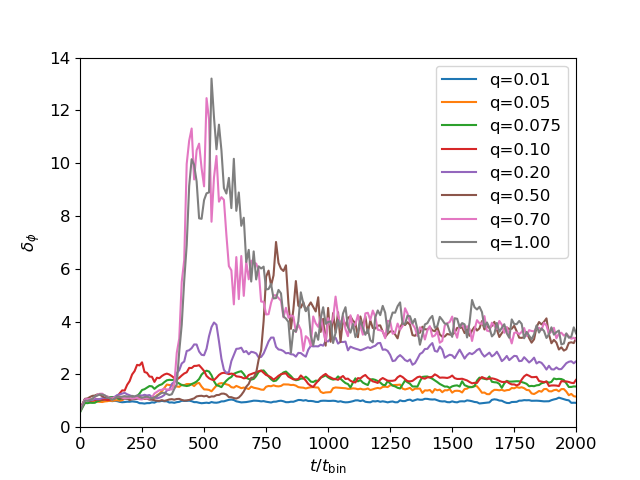}
	\includegraphics[width=\columnwidth]{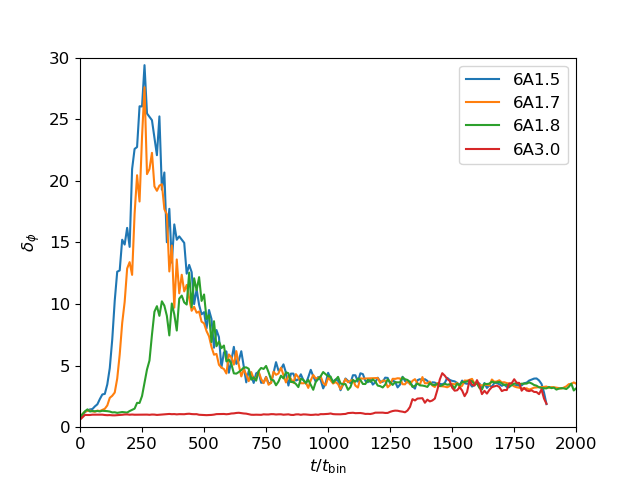}
    \caption{Density azimuthal contrast ratio of as a function of time for simulations ``A'' with different mass ratios (left panel) and simulations 6A1.5, 6A1.7, 6A1.8 and 6A3.0 (right panel) (see Tab. \ref{Tab:sims}). We note that when the cavity size grows while the disc becomes eccentric, for $q\geq 0.2$ the disc develops a pronounced azimuthal asymmetry, that progressively decays  to a density contrast of $\sim 4$ after $\sim 1000$ binary orbits, consistent with that expected from an eccentric ``traffic jam''. Larger initial inner disc radii $R_{\rm in}$ postpone the time the disc transitions eccentric configuration and reduces the maximum contrast ratio the over-density can achieve.}
    \label{fig:cratio}
\end{figure*}

As previously mentioned, when the transition to an ``eccentric'' configuration occurs, the disc develops a prominent azimuthal over-density at the cavity edge with the shape of a horseshoe. This feature can be seen in all our simulations with mass ratio $q> 0.2$. Its initial contrast ratio grows with the binary mass ratio, as shown in Fig. \ref{fig:cratio} where we evaluate the contrast ratio $\delta_\phi$ of an over-dense feature by averaging the surface density in the surroundings of its maximum value and comparing it with the value at the opposite side of the cavity.
Each data point for the value of $\delta_\phi$ in Fig. \ref{fig:cratio} is obtained performing a moving average over a window of 5 binary orbits. This makes the plot significantly less ``noisy'', but it smooths away fluctuations in the contrast ratio that may occur on timescales shorter than 10 binary orbits.

\begin{figure*}
	\includegraphics[width=\textwidth]{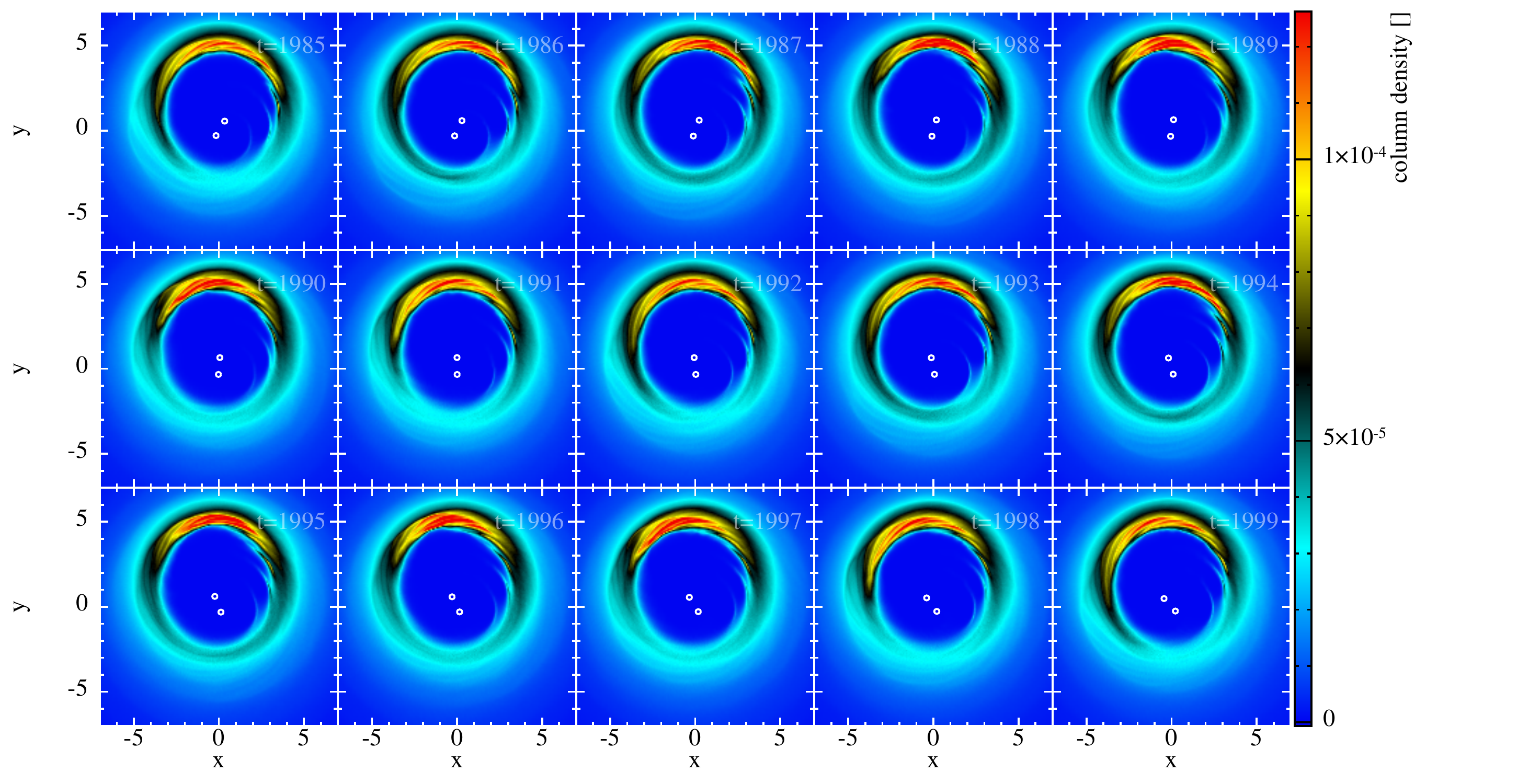}\\
	\caption{Surface density map of simulation 6A for times $t=1985 \textrm{--} 1999 \,t_{\rm orb}$, using a linear colour scale. The plot is meant to show that, besides the ``eccentric'' over-dense feature at the cavity apocentre, a periodic ($\approx 7\textrm{--}8\,t_{\rm orb}$) variation of the density at cavity pericentre still occurs at the end of the simulation. The reader will notice that, besides the higher contrast non-orbiting feature at the cavity apocentre (North of each snapshot), at the cavity pericentre (South) the surface density varies by a factor $\approx 1.5\textrm {--}2$ every $\approx 7\textrm{--}8\,t_{\rm orb}$. This suggests that a low contrast over-density ($\delta_\phi\sim 1.5\textrm{--}2$) is still orbiting, co-moving with the flow, at the edge of the cavity. See also Fig. \ref{fig:periodogram}.} \label{fig:snapshotsHorse}
\end{figure*}

\begin{figure}
	\includegraphics[trim={0cm 0.cm 1cm 1cm},clip,width=\columnwidth]{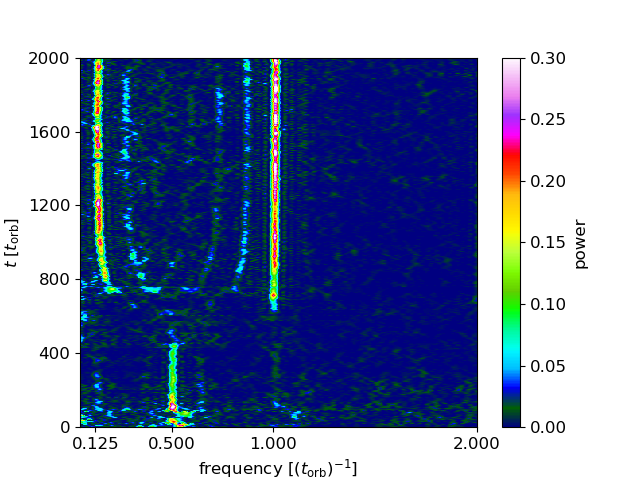}\\
	\caption{Periodogram of the accretion rate $\dot M$ on to the binary (frequency is on the x-axis and powers are coded with different colours) at different times (y-axis). The x-axis report frequencies in units of $(t_{\rm orb})^{-1}$, so that the vertical line centred at frequency $t^{-1}=0.5\,(t_{\rm orb})^{-1}$ for the first 500 binary orbits and $t^{-1}=1\,(t_{\rm orb})^{-1}$ from $t\gtrsim 800\, t_{\rm orb}$ indicates that the binary shows a modulation of the accretion rate once every two binary orbits and once every orbit, respectively. On the top of that, a slower modulation with frequency $t^{-1}=0.12\textrm{--}0.13\,(t_{\rm orb})^{-1}$ appears as soon as the orbiting over-dense lump forms ($t\gtrsim 800\,t_{\rm orb}$). Such an accretion feature is linked to the presence the orbiting over-density which makes a close passage at the cavity pericentre every $7\textrm{--}8\, t_{\rm orb}$ (as also shown in Fig. \ref{fig:snapshotsHorse}). The presence of such accretion feature at late times implies that an orbiting over-dense lump of material is still present at the end of the simulation.} \label{fig:periodogram}
\end{figure}

In order to better capture the orbiting/non-orbiting nature of such features, we introduce also two additional figures. First, Fig. \ref{fig:snapshotsHorse} shows 15 snapshots towards the end of simulation 6A ($q=0.5$, times shown are $t= 1985\textrm{--}2000 t_{\rm orb}$, one per binary orbit). Second, Fig. \ref{fig:periodogram} shows a Lomb-Scargle periodogram of the accretion rate on to the binary throughout the length of simulation 6A. Frequencies are on the x-axis (in units of $(t_{\rm orb})^{-1}$), colours code the powers of different frequencies and times are on the y-axis. Horizontal lines ($t={\rm const}$) in this plot represent the periodogram of the accretion rate on a window of 30 binary orbits, at a fixed time of the simulation.

From now on, when relevant, we will distinguish among azimuthal over-dense features referring to them as ``orbiting over-dense lumps'', when the feature moves with Keplerian motion at the edge of the cavity, and ``eccentric traffic jam'', for non orbiting features. We discuss the formation mechanism of these features in detail in Sec. \ref{sec:horseform}.

\subsection{Results for Different Disc Parameters}
Here we present a second set of simulations we performed for a fixed mass ratio $q=0.2$ while varying some of the disc parameters.
Right panels in Fig. \ref{fig:cav_A} and \ref{fig:ecc_A} show the time evolution of the cavity size (left panel) and ``total'' disc eccentricity (right panel) using Eq. (\ref{eq:efromAMD2}) for the simulations: 5A, 5C, 5E, 5Z, 5N, 5O, 5P, 5H and 5A3.0 in Tab. \ref{Tab:sims}. We first note that interestingly the case 5A3.0 suggests that the transition to the ``eccentric'' disc configuration takes place only when the still circular cavity edge reaches a minimum separation from the binary. Indeed, simulation 5A3.0, being initialised with a larger cavity ($R_{\rm in}=3$), shows a delay in the growth of the disc eccentricity, probably due to the need for the disc to spread viscously until it reaches some resonant location. The slope of the density profile (5O and 5P), the disc mass (5H), and small changes in the disc viscosity (5Z) are not causing significant differences in the evolution of the disc eccentricity. An increased disc thickness (sim 5C), besides opposing the opening of a cavity due to the stronger pressure gradient at the cavity edge, increases the disc viscosity $\nu$, since it is parametrised using the \citet{shakura1973} prescription. This provides a faster spread of the disc disc towards the resonant location and transition to the disc eccentric configuration at earlier times, even though the maximum disc eccentricity is lower than in the reference case. Explicitly increasing the disc viscosity through the $\alpha_{\rm ss}$ parameter (sim 5E) produces the same effect. Consistent with this scenario, a reduction in the disc thickness (sim 5N) shows that the disc transition to the eccentric configuration occurs at later times. In this last case the final value of the disc eccentricity is higher than in the reference case 5A.
The right panel in Fig. \ref{fig:ebin} shows how different disc parameters affect the evolution of the binary eccentricity.

In a similar fashion, we also varied the disc parameters while fixing the mass ratio $q=0.5$ (simulations 6C, 6Z, 6E). Simulations 5A3.0, 6A1.5, 6A1.7, 6A1.8 and 6A3.0 all share the same disc properties of ``A'' discs, with the only difference that their initial inner truncation radius is set to $R_{\rm in}=\{1.5,1.7,1.8,3.0\}\,a_{\rm bin}$ according to their reference label, as outlined in Table \ref{Tab:sims}. The surface density evolution from these simulations appear in Fig. \ref{fig:6SmallLarge}, they will be discussed further in Sec. \ref{sec:horseform}.

\section{Discussion}\label{sec:discussion}

The results presented in the previous sections hint at a number of interesting features in the evolution of eccentric discs. We note that items i-iii below have been previously discussed in the literature, items iv-viii, to our knowledge, did not receive the same attention and will be subject of a deeper discussion.

\begin{figure*}
	\includegraphics[width=\columnwidth]{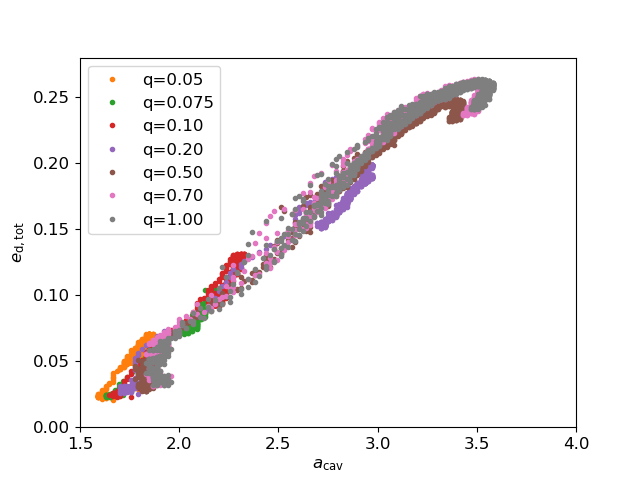}
	\includegraphics[width=\columnwidth]{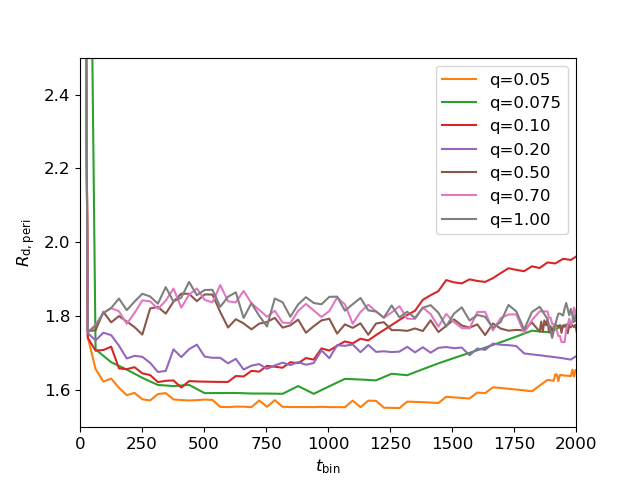}
	\caption{Left panel: Correlation between disc eccentricity (y-axis) and cavity semi-major axis (x-axis) throughout the entire length of the simulation for different ``A'' discs. Right panel: the cavity pericentre radius as a function of time, computed using $R_{\rm d,peri}=a_{\rm cav}(1-e_{\rm d,tot})$ for different ``A'' discs. As in the left panel of Fig. \ref{fig:cav_A} we deliberately omit the case $q=0.01$.
	} \label{fig:correacav}
\end{figure*}

\begin{enumerate}
\item All discs with sufficiently large binary mass ratios (which here appears to be $q\gtrsim 0.05$) become eccentric, consistently with what previously found \citep{dorazio2016,ragusa2017,munoz2020b}. Previous studies have shown that binaries with mass ratios $q<0.05$ may excite the eccentricity of the circumbinary disc \citep{dangelo2006,kley2006,teyssandier2017}.  Nevertheless, in this work we mainly refer to the abrupt growth of the disc eccentricity that occurs in most of our simulations after an initial ``circular phase'' that lasts for $\sim 400-700$ orbits (see Fig. \ref{fig:simA} and top panel of \ref{fig:pericentre_A} -- see also \ref{fig:snapshots} below).

\item Eccentric discs undergo rigid longitude of pericentre precession (the pericentre of the eccentric disc orbits remains aligned throughout the entire disc, see bottom panel of Fig. \ref{fig:pericentre_A}).
The precession rate of the disc is independent from that of the binary, which precesses at a much slower rate (Fig. \ref{fig:pericentre_A2}). This had been previously found numerically \citep{macfadyen2008,miranda2017,thun2017,ragusa2018} and discussed theoretically by \citet{teyssandier2016}, and recently by \citet{munoz2020b}. The physical explanation of the origin of this pericentre alignment is that the clustering of eccentric orbits at the apocentre is ``pinching'' \citep{dermott1980}\footnote{We note that in \citep{dermott1980} the ``pinching'' occurs at the pericentre as the eccentricity profile has a positive gradient.} together all the elliptic orbits, preventing them from precessing differentially.

\item As noted above, a prominent orbiting over-dense lump develops for mass ratios $q>0.2$ that is co-moving with the gas (i.e. not an eccentric ``traffic jam'') as shown in Fig. \ref{fig:snapshots}.
This feature has been observed in previous works \citep[e.g.][]{farris2014,miranda2017,ragusa2017} and is referred to as ``over-dense lump'' or ``horseshoe feature'' in the black hole and protoplanetary disc community, respectively. A discussion about its formation mechanism has been provided by \citet{shi2012}.  However, many aspects regarding formation and evolution of such features remain unclear -- see Sec. \ref{sec:horseform} for further discussion.

\item In all cases the binary gains a small amount of eccentricity before entering the ``eccentric cavity'' phase -- $e_{\rm bin}\sim 0.001-0.007$ in most cases. Two cases, which respectively used a more massive and a thinner disc than the reference case (simulations 5H, larger disc mass, and 5N, lower disc thickness, in Table \ref{Tab:sims}), rapidly reach $e_{\rm bin}\sim 0.01$ and keep growing. This behaviour is consistent with previous studies \citep{dunhill2013,ragusa2018}, which found that a larger disc-to-binary mass ratio $q_{\rm d}=M_{\rm d}/(M_1+M_2)$ (sim 5H, $q_{\rm d}=0.01$) leads to a larger binary eccentricity. The higher value of the binary eccentricity associated to a thinner disc (sim 5N, $H/R=0.03$) is consistent with a reduction of the resonance width for lower values of $H/R$, which provides a stronger Lindblad torque on the binary \citep{meyerV1987}. See Eq.s (21) and (22) in \citet{goldreich2003} for the dependence of binary torque on both disc mass and resonance width.

\item The duration of the initial phase during which the disc remains circular depends on the initial radius of the disc, suggesting that the disc spreads viscously and then encounters resonances. Our results suggest that the resonances located at the 1:2 frequency commensurability play a role in explaining the evolution of the disc eccentricity we observe, as previously suggested by \citet{dangelo2006}. We will discuss further about the role of different resonances (or alternative non-resonant mechanisms) for the growth of the cavity eccentricity in Sec. \ref{sec:horseform}. We will see that it is hard to interpret the results within the theoretical framework currently available in the literature, posing the basis for possible future developments of the theory.

\item In all the simulations, the disc eccentricity stops growing when it reaches a maximum value. For $q>0.5$, the disc eccentricity appears to saturate at a maximum value which is independent of the binary mass ratio -- inner edge of the cavity $e_{\rm d}(a_{\rm cav})\sim0.5$, ``total'' eccentricity $e_{\rm d,tot}\sim0.25$. For $q<0.5$, this maximum value scales with the disc viscosity (see Fig. \ref{fig:ecc_A}). We discuss this further in Sec. \ref{sec:horseform}.

\item When discs become eccentric, they appear to have larger cavities than when they are circular: the cavity semi-major axis becomes up to 3.5 times the binary separation, whereas during the start of the simulations
the inner edges of the disc all remain at
$\sim 2$ binary separation for $\sim 400$ binary orbits (see Fig. \ref{fig:simA} and left panel of Fig. \ref{fig:cav_A}). There is a strong correlation between the cavity eccentricity and its size, as shown in the left panel of Fig. \ref{fig:correacav}. We will discuss this aspect further in Sec. \ref{sec:resostrength}.

\item  The radius of the cavity pericentre, the evolution of which is shown in the right panel of Fig. \ref{fig:correacav}, remains approximately constant at $R_{\rm d,peri}=a_{\rm cav}(1-e_{\rm d,tot})\lesssim 2$ throughout the simulation. This is consistent with the correlation found between $a_{\rm cav}$ and $e_{\rm d,tot}$. This suggests that the minimum separation of gas particles from the binary is fixed for a given mass ratio, and growth of the disc eccentricity (at fixed pericentre) therefore results in corresponding growth of the cavity semi-major axis -- see Sec. \ref{sec:resostrength} for further discussion.
\end{enumerate}

In the following sections we interpret these results within the existing theoretical framework. We also speculate about possible interpretations of some results that cannot be explained with our current understanding of resonant and non-resonant binary-disc interaction, hinting at the direction that further theoretical studies should take to confirm the interpretation of our numerical results.

\subsection{Evolution of the cavity eccentricity}\label{sec:evodiscecc}

In Sec. \ref{sec:resointro} we discussed the role resonant interaction is expected to play for the evolution of both binary and disc eccentricity. Non-resonant mechanisms might also play a role.
We identify four possible sources of the eccentricity growth.

\begin{enumerate}
\item Resonance $(m,l)=(3,2)$ ELR located at the 1:2 binary-to-disc orbital frequency commensurability ($R_{\rm L}=1.59\,a_{\rm bin}$). This resonance is expected to pump the disc eccentricity \citep{goldreich2003}; its role in the evolution of the disc eccentricity has been previously discussed by \citet{dangelo2006}. This resonance requires the binary eccentricity to be $e_{\rm bin}\neq0$ to be effective, which, despite small values of binary eccentricity are excited, is the case for our simulations.

\item At the same location as the $\{m,l\}=\{3,2\}$ ELR ($R_{\rm L}=1.59\,a_{\rm bin}$), lies the $(m,l)=(1,1)$ OCLR, which is expected to pump the disc eccentricity as well. This resonance is effective also for circular binaries. We note that OCLRs are the only available resonances that can increase the disc and binary eccentricity if the binary is fixed on a circular orbit \citep{macfadyen2008}.

\item Resonance $\{m,l\}=\{2,1\}$ ELR at the 1:3 frequency commensurability ($R_{2,1}=2.08\times a_{\rm bin}$) has also been previously suggested to play a role in eccentricity evolution by \citet{papaloizou2001}.

\item  The lack of stable closed orbits around Lagrange points L4 and L5 for binary mass ratios $q>0.04$ has been discussed to be possibly causing the growth of the disc eccentricity \citep{dorazio2016}; this mechanism is non-resonant.

\item Impact of gaseous streams from the disc cavity pericentre hitting the opposite edge of the cavity wall \citep{shi2012,dorazio2013}. This mechanism is non-resonant.
\end{enumerate}

The disc eccentricity has been shown to increase exponentially when a small eccentricity seed in the disc is present (see Eq. (14) in \citealp{teyssandier2016}). Thus, as soon as the disc and/or the binary have a small fluctuation in their orbital eccentricity, if the disc covers an OCLR/ELR location, the eccentricity increases rapidly until some other physical mechanism limits its growth. Non-resonant growth of the disc eccentricity has not been quantitatively discussed in the literature. However, in essence, as soon as the gas spreads towards the co-orbital region it will be forced to move on non-closed orbits, perturbing the circularity of the cavity. As for resonant mechanisms, non-resonant eccentricity growth stops when eccentricity damping through some secondary mechanism becomes dominant.

\begin{figure*}
	\includegraphics[width=\textwidth]{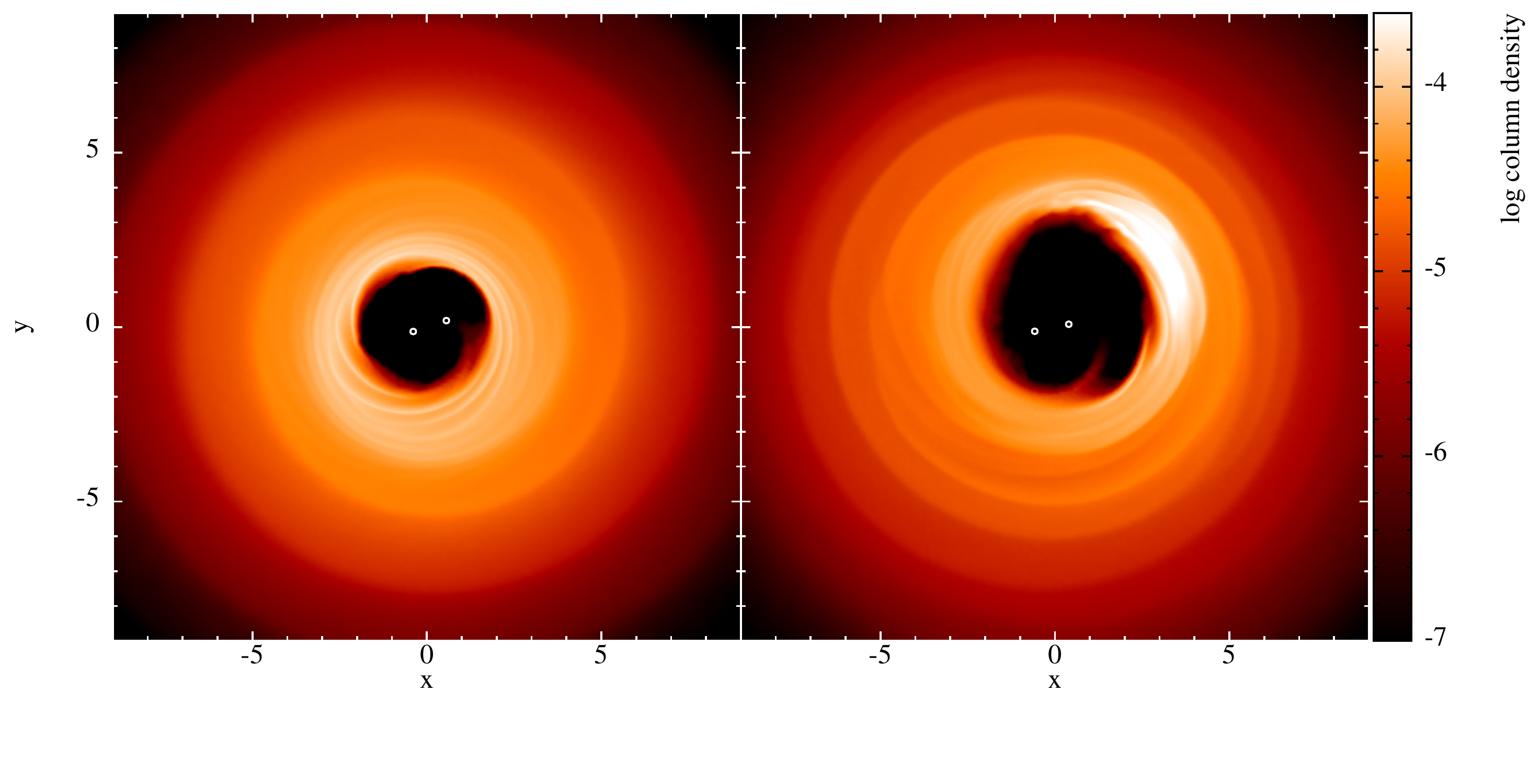}\\
	\caption{ Snapshots of the case 6A ($q=0.7$) after $t=330\, t_{\rm orb}$ (left panel) and $t=450 \,t_{\rm orb}$ (right panel). A Three lobed cavity (left panel) marks transition from a ``small'' circular cavity to a ``large'' eccentric cavity.} \label{fig:snapshots}
\end{figure*}

As soon as the simulation starts, the disc viscously spreads inward from its initial radius, without growing its eccentricity at all.
Then a three-lobed structure appears (see Fig. \ref{fig:snapshots}). Immediately after the appearance of this three-lobed structure, the disc eccentricity rapidly increases toward its maximum value.

Our simulations were started with $R_{\rm in}=2$ in order to deliberately cover the 1:3, $\{m,l\}=\{2,1\}$, binary-disc orbital frequency commensurability, which is located at $R_{\rm L,21}=2.08\, a_{\rm bin}$. Given the delay in the growth of the eccentricity, we can exclude this resonance as being the main contributor to the abrupt growth. Resonances $\{m,l\}=\{3,2\}$ and $\{m,l\}=\{1,1\}$, located at the 1:2 binary-disc orbital frequency commensurability ($R_{\rm L,32}=1.59\, a_{\rm bin}$) appear to be better candidates \citep[see also][]{dangelo2006,macfadyen2008,miranda2017}. The right panel of Fig. \ref{fig:cratio} shows that reducing the disc inner radius $R_{\rm in}$ at the beginning of the simulation moves forward the abrupt growth of the disc eccentricity and cavity size (producing the high contrast ratio showed in that plot); the growth of eccentricity starts immediately when $R_{\rm in}=1.5<R_{\rm L,32}$ for simulation 6A1.5. Resonance $\{m,l\}=\{1,1\}$, being a circular resonance, appears to be the strongest resonance among those proposed. However, the formation of the three-lobed structure suggests an $m=3$ resonance is effective, implying that also $\{m,l\}=\{3,2\}$ ELR resonance might be playing a role, despite the binary eccentricity being small  $e_{\rm bin}\lesssim 0.01$. Non-resonant eccentricity growth cannot be excluded, but it is hard to justify the $m=3$ spiral in that framework.

We note that for binaries with mass ratio $q=1$ the OCLR resonance $\{m,l\}=\{1,1\}$ is not effective. \citet{macfadyen2008} showed that the OCLR resonance $\{m,l\}=\{2,2\}$ located ad the 3:2 binary-to-disc frequency commensurability ($R_{\rm L,22}=1.31\, a_{\rm bin}$) is effective for the growth of the eccentricity for this specific case. However, it is not clear whether in our simulations some material reaches that separation before the abrupt disc eccentricity growth starts. More generally, our simulation with $q=1$ (simulation 8A) does not show sufficient evolution of the binary eccentricity for ELRs to be effective, making the growth of the disc eccentricity for the $q=1$ case hard to justify within our current understanding of resonances.
In Sec. \ref{sec:resostrength} we will speculate about the possible growth of the intensity of ELRs to solve this and other issues that will arise when discussing the cavity size.

The evolution of the disc eccentricity is ruled by competing effects that damp or pump the eccentricity. We mentioned in Sec. \ref{sec:resointro} that co-rotation resonances damp the eccentricity, but they are expected to saturate and lose their circularising effect. Viscous dissipation and eccentric orbit intersection, which occurs when the disc eccentricity gradient is sufficiently steep to satisfy the following criterion \citep{dermott1980}
\begin{equation}
a\left(\frac{{de}}{da}\right)\gtrsim 1,\label{eq:orbitcross}
\end{equation}
then become the main mechanisms acting to damp the disc eccentricity.
Nested elliptical orbits with eccentricity decreasing with radius are expected to be subject to friction which becomes stronger depending on the eccentricity gradient.

Viscous dissipation occurs when the eccentricity pumping effect is progressively balanced by the damping effect provided by viscosity. The criterion for orbit intersection is instead a physical limit, beyond which strong shocks rapidly damp the eccentricity. This explains the maximum values the disc eccentricity can reach in our simulations. With reference to Fig. \ref{fig:ecc_A}, in our reference simulations (``A'' simulations in Tab. \ref{Tab:sims}) cases with $q<0.5$ progressively grow their eccentricity until viscous circularisation balances the effect of the ELRs. For larger mass ratios ($q\geq 0.5$), ELRs can in principle pump more eccentricity in the disc, but orbit intersection prevents its further growth, causing the eccentricity to saturate to the same maximum value for all $q\geq0.5$. Fig. \ref{fig:ecrit}
shows a comparison between the values of Eq. (\ref{eq:orbitcross}) throughout the disc. We see clearly that the region at the edge of the cavity fully satisfies the criterion for orbit intersection for $q=0.7$ (sim 7A), but not for $q=0.2$ (sim 5N).
Despite reaching very similar maximum values of disc eccentricity, the case 5N, with a thinner disc, has a lower value of disc viscosity and thus the viscous damping is weakened, allowing the eccentricity to grow to larger values, with respect to the reference $q=0.2$ case (sim 5A). However the steepness of the eccentricity profile does not appear to be high enough to provide orbit intersection.

Orbits do not intersect for shallow eccentricity profiles for any value of the eccentricity (see Fig. \ref{fig:ecrit}).
The maximum value the eccentricity can reach is thus determined by two mechanisms. When the right resonances are excited, the binary pushes the disc to become eccentric, while the viscous dissipation in the disc tends to circularize the orbits. When these two processes balance, the eccentricity stops evolving.

\begin{figure*}
	\includegraphics[width=0.49\textwidth]{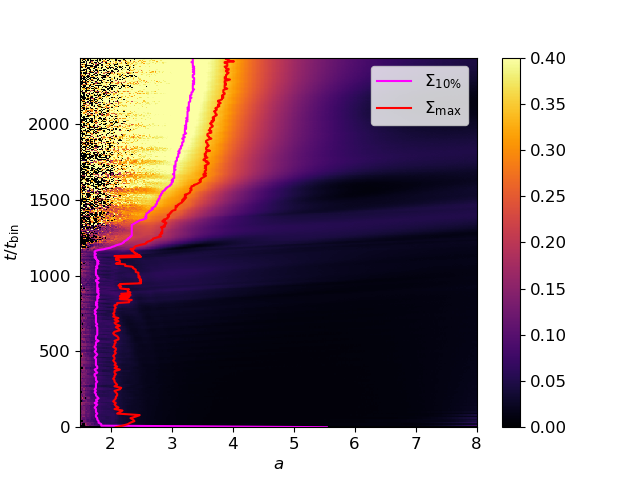}	\includegraphics[width=0.49\textwidth]{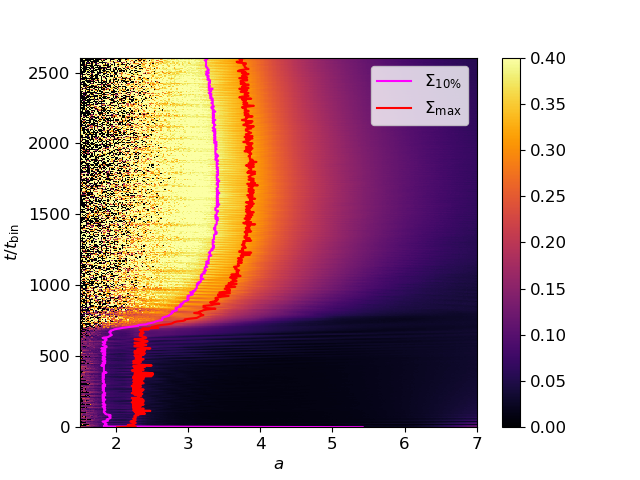}
	\caption{Disc eccentricity as a function of time (y-axis) and semimajor axis (x-axis) for the two cases 5N (left-hand panel) and 7A (right-hand panel); see Table \ref{Tab:sims}. A higher eccentricity can be achieved at the cavity edge of case 5N (between the purple and red lines) than in case 7A, since orbits are not intersecting (see also Fig. \ref{fig:ecrit}).}\label{fig:eAvstR}
\end{figure*}

\begin{figure*}
	\includegraphics[width=0.49\textwidth]{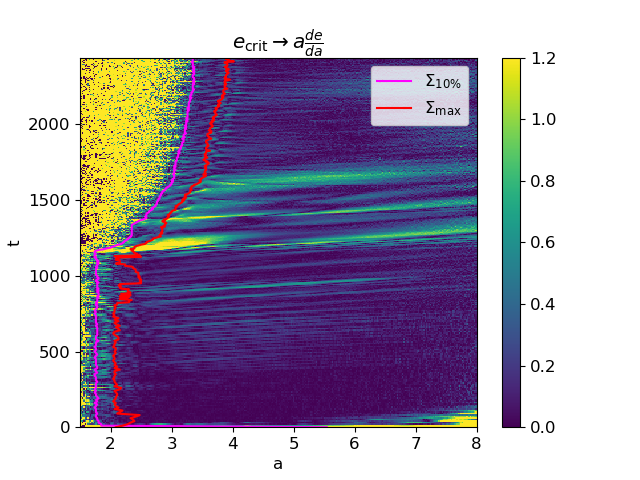}
	\includegraphics[width=0.49\textwidth]{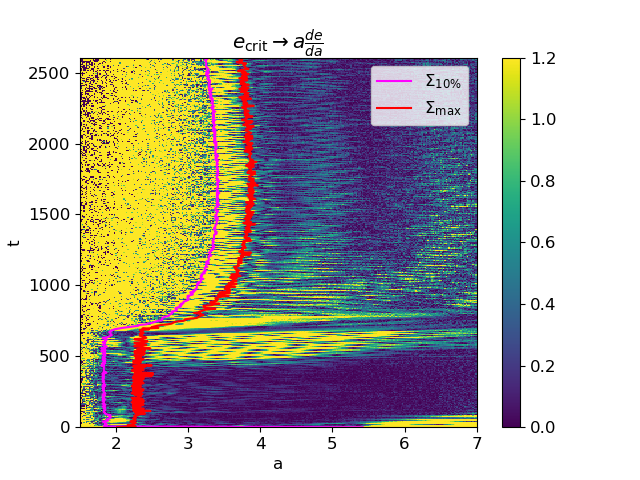}
	\caption{Colour plot of the quantity in equation (\ref{eq:orbitcross}) as a function of time (y-axis) and semimajor axis (x-axis) for the cases 5N (left-hand panel) and 7A (right-hand panel), as in Table \ref{Tab:sims}. When $a\, de/da > 1$ eccentric orbits are expected to cross, suppressing further eccentricity growth in the disc, as shown in Fig. \ref{fig:eAvstR} where the case 5N reaches a larger value of maximum eccentricity at the cavity edge. }\label{fig:ecrit}
\end{figure*}

\subsection{Cavity size: strengthening of Eccentric Resonances or non-Resonant Truncation?}\label{sec:resostrength}

If small binary eccentricities as those we observe in our simulations can in principle activate ELRs, causing them to produce typical density features -- as the m=3 spiral right before the onset of the eccentricity growth -- it is very hard to believe they have sufficient strength to truncate the disc.

In our simulations we observe an initial growth of the binary eccentricity (see Fig. \ref{fig:ebin}), which could in principle cause the strength of ELRs to grow.

\begin{figure}
	\includegraphics[width=\columnwidth]{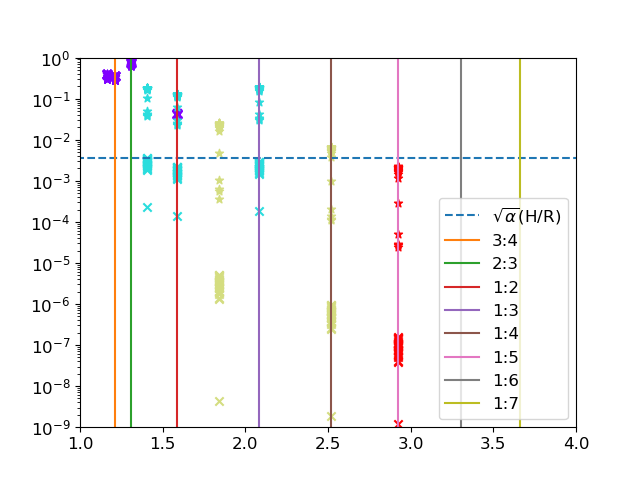}
	\caption{Intensity of individual resonances using the r.h.s. of Eq. (16) by \citet{artymowicz1994} (y-axis)  vs their location in the disc (x-axis). The dashed line represents the viscous threshold in their intensity in order to create a depletion in the disc surface density. Each intensity is computed at different times using the orbital properties of the binary (crosses). Star markers are instead computed using the disc eccentricity. We use different marker colours to indicate the order of ELRs: namely, purple are first order ($m=l$), cyan second order ($m=l+1$), olive third order ($m=l+2$), coral fourth order ($m=l+3$). We note that this is not meant to prove our claims, but it simply shows that if the disc eccentricity affects the intensity of resonances in the same way as the binary, a growth of the cavity size is reasonable to occur up to the 1:5 resonant location.} \label{fig:resostrength}
\end{figure}

However, the values of binary orbital eccentricity that are excited in our simulations are in most cases $e_{\rm bin}<0.01$. Our current understanding of ELRs tells us that for such small values of binary eccentricity, resonances cannot overcome the viscous forces in the disc to open a large cavity \citep{artymowicz1994}. We show this in Fig. \ref{fig:resostrength}, where we provide an estimate of the intrinsic strength of ELRs whose location in the disc is consistent with the size of the cavity, and then compare it to a criterion for a cavity to be opened by that resonance \citep{artymowicz1994}.

OCLRs cannot be invoked to explain disc truncation at radii $R>R_{\rm L,11}=1.59\,a_{\rm bin}$, being the $\{m,l\}=\{1,1\}$ the outermost circular Lindblad resonance. Given the low strength of ELRs in our simulations, we cannot invoke resonant truncation to explain cavities as large as $a_{\rm cav}\approx 4\,a_{\rm bin}$ shown in Fig \ref{fig:cav_A}.

We here speculate about two possible scenarios that can be responsible of the depletion of such large cavities.

First, the intrinsic strength of ELRs also depends on the disc eccentricity, instead of being exclusively related to the binary one. This possibility would set the basics for a new physical mechanism producing the unstable growth of cavity size, which relates with the disc eccentricity since the the higher its value is, the stronger the outer ELRs are. This speculative scenario is supported by calculations that use fixed binaries (i.e., not allowed to change their orbital parameters, as we allow here) showing an evolution of the cavity structure in terms of size and eccentricity beyond the location of the outermost circular Lindblad resonance (e.g. \citealp{kley2006,shi2012,farris2014,dorazio2016}), implying that having $e_{\rm bin}\neq 0$, is not a fundamental requirement in order to activate ELRs.

Second, a non-resonant mechanism sets the cavity size \citep{papaloizou1977,rudak1981,pichardo2005,pichardo2008}. In this interpretation disc truncation takes place at the innermost separation from the binary where gas orbits are not anymore ``invariant loops'' \citep{pichardo2005,pichardo2008}, implying that gas particle orbits are ``intersecting'', dissipating the orbital energy and clearing the cavity region -- which is analogous to what happens when the eccentricity gradient exceeds the ``orbit intersection'' threshold (see above Eq. (\ref{eq:orbitcross}) in Sec. \ref{sec:evodiscecc}).

Previous studies which considered the dependence of the truncation radius of the cavity on the binary properties assumed circular orbits in the disc \citep{pichardo2005,pichardo2008}.
We speculate that non-intersecting orbits of the gas are expected up to a minimum separation from the binary, and this sets the pericentre radius of the cavity edge. This is supported by the behaviour of the cavity pericentre, plotted in the right panel Fig. \ref{fig:correacav}: where the pericentre radius depends on the binary eccentricity and mass ratio, but remains roughly constant for all the mass ratios throughout the simulation.

If the pericentre radius $R_{\rm d,peri}$ of the cavity is fixed non-resonantly, when the disc eccentricity grows, it will cause the growth of the cavity semi-major axis.
The correlation between the cavity size and disc eccentricity shown in Fig. \ref{fig:correacav} supports this scenario. The relationship between the disc eccentricity and innermost non-intersecting orbit has not been established yet, and will be subject to future studies.

Completing this second scenario, we note that if the non-resonant eccentricity growth scenario introduced in Sec. \ref{sec:evodiscecc} is effective, the mechanism we describe would be completely non-resonant.

However, we note that we are not able to verify this interpretation without substantial further development of the theoretical framework of resonant and non-resonant binary-disc interaction for eccentric discs. Indeed, we are not aware of any previous work suggesting or directly studying these effects.

\subsection{Formation of the azimuthal over-density}\label{sec:horseform}

One of the most interesting features that arises from these simulations is the formation of a well defined azimuthal asymmetry in the density field (see right panel of Fig. \ref{fig:snapshots}) in all simulations with binary mass ratio $q>0.2$. This feature is often referred to as a ``horseshoe'' in the protoplanetary disc community, due to its shape, or ``over-dense lump'' in the black hole community. It has been seen in a number of previous studies \citep{shi2012,farris2014,ragusa2016,ragusa2017,miranda2017, calcino2019}. In order to evaluate the intensity of the asymmetry we define the contrast ratio $\delta_\phi$ as the ratio between the density in the azimuthal feature, and the density at the opposite side of the cavity.

If the disc eccentricity profile has a negative radial gradient, it is expected to form an non-orbiting azimuthal over-density with contrast ratio $\delta_\phi\approx 3\textrm{-- }4$. This is referred to as the ``eccentric feature'' or ``traffic jam'', and caused by the clustering of orbits at their apocentres \citep{ataiee2013,teyssandier2016,thun2017}.
This feature (we refer to it as ``eccentric traffic jam'') is fixed at the apocentre of the cavity, and moves only because of the precession of the longitude of pericentre of the cavity.

The over-density visible in Fig. \ref{fig:snapshots} (we refer to it as ``orbiting over-dense lump'') not only moves around the edge of the cavity with Keplerian motion, but reaches a contrast ratio $\delta_\phi \geq 10$, larger than the typical values in ``traffic jams''.

Tidal streams being thrown from the cavity pericentre against its opposite edge have been discussed to cause the formation of the feature \citep{shi2012,dorazio2013}. Consistent with this picture, from our work it emerges that one of the key elements for the formation of a high contrast ratio over-density is the fast outward motion of the gas when the cavity progressively increases in size. Moving outwards, the gas first produces an over-dense ring of material, which then evolves into an azimuthal structure.

\citet{ragusa2017} found a threshold value for the formation of high contrast over-densities of $q>0.05$. Simulations in that work used discs that initially extended up to $R_{\rm in}=1.5<R_{\rm L,11}$, making that result consistent with what we found in this paper.
Even though no whirling motion is present in orbiting over-dense lumps, \citet{hammer2017} noted that RWI vortices are weaker when planets inducing them appear in the simulation slowly increasing their mass; similarly, this leads to a slower buildup of material at the gap/cavity edge.

\subsubsection{Evolution and life expectation of the azimuthal over-dense feature}\label{sec:od_lifetime}

This qualitative picture described above appears to be confirmed by the evolution of the contrast ratio. Fig. \ref{fig:cratio} shows the contrast ratio of the azimuthal over-density, $\delta_\phi$, as a function of time for our ``A'' reference cases and for four different choices of initial $R_{\rm in}$ of the simulation (simulations 6A1.5, 6A1.7, 6A1.8 and 6A3.0).

The contrast ratio of the azimuthal over-density grows when the cavity size starts growing, reaching a peak when the cavity size reaches its maximum value. When the material stops moving outward, the over-density stops growing.
After its peak, the contrast ratio of the over-density progressively decreases -- probably due to viscous dissipation -- until it appears to stabilise at a value of $\approx 4$ for $q\geq 0.2$, which are characterised by the maximum eccentricity gradient being fixed by orbit-crossing limit, Eq. (\ref{eq:orbitcross}). This final value of the contrast ratio corresponds to the contrast ratio of the slowly precessing ``eccentric traffic jam'' structure that forms at the cavity apocentre following the growth of the disc eccentricity.

Initializing the disc with a larger inner radius $R_{\rm in}$ leads to later growth of the cavity size and eccentricity (Fig. \ref{fig:6SmallLarge}). Since the amount of material that viscously spreads inward is less than that present in a simulation starting with the disc extending to smaller inner radii, the amount of material pushed outward when the cavity becomes eccentric and increases its size is lower, resulting a less pronounced over-dense feature. When starting the simulation with $R_{\rm in}=3$ (simulation 6A3.0) the system seem to directly form an ``eccentric traffic jam'' feature.

Despite the ``eccentric traffic jam'' becoming the dominant over-density at late times, a feature with a contrast ratio of $\sim 1.5\textrm{--}2$ keeps orbiting at the cavity edge, as shown in Fig. \ref{fig:snapshotsHorse}. This result is consistent with what previously found by \citet{miranda2017}, where the authors found that an orbiting over-density with contrast $\delta_\phi\sim 2\textrm{--}3$ is found to be present after 6000 binary orbits, with a viscosity 10 times higher than the one used in the present work\footnote{\citet{miranda2017} used $\alpha_{\rm ss}=0.05$ for the aforementioned simulation.}.
Such feature makes a close passage to the binary when it reaches the cavity pericentre, i.e. every $\sim 7\textrm{--}8$ binary orbits. This boosts the accretion rate with the same periodicity (see Fig. \ref{fig:periodogram}) as previously found in a number of work \citep[e.g.][]{cuadra2009,dorazio2013,farris2014,ragusa2016,miranda2017}.

We note that in our analysis (not shown here to reduce the paper length) also simulation 6A3.0, where the contrast ratio never exceed $\delta_\phi=4$, shows a low contrast ratio ($\delta_\phi\sim 1.5$) orbiting over-dense feature that produces a modulation of the accretion rate of 7--8 binary orbits.

Even though the orbiting over-dense lump maintains a contrast ratio $\delta_\phi\gtrsim 10$ for a limited number of orbits ($\sim 1000\,t_{\rm orb}$), it survives with a low contrast ($\delta_\phi\approx 1.5\textrm{--}2$) for longer timescales. This result does not depend on the previous history of the evolution of $\delta_\phi$, as mentioned before for simulation 6A3.0, where the contrast ratio never exceeded $\delta_\phi=4$.

These results may explain azimuthal over-dense features observed in dust thermal emission in protoplanetary discs (see Sec. \ref{sec:observ} for further discussion).

\subsection{Effects of Different Disc Parameters}

The right panels of Fig. \ref{fig:cav_A} and \ref{fig:ecc_A} show how disc parameters affect the eccentricity and cavity size. The relevant parameters for the long time-scale evolution appear to be $\alpha$, $H/R$, and $q$. Apart from some effects on the initial evolution all the simulations evolve towards the same long time-scale behaviour apart from simulation 5N (larger cavity, higher final disc eccentricity, $H/R=0.03$ i.e. smaller than the ``A'' reference case), simulation 5E (smaller cavity, lower final disc eccentricity, $\alpha=0.1$ i.e. larger than the ``A'' reference case), and simulation 5C (smaller cavity, lower final disc eccentricity, $H/R=0.1$ i.e. larger than the ``A'' reference case). In the \citet{shakura1973} prescription the viscosity is $\nu \propto \alpha (H/R)^2$, so the behaviour of simulations 5N, 5E and 5C can be mainly attributed to the resulting differences in the disc viscosity. However, as previously mentioned in Sec. \ref{sec:discussion}, the width of these resonances scales as $(H/R)^{2/3}$ \citep{meyerV1987,teyssandier2016}, so the disc thickness may also play a role in determining the evolution of the binary eccentricity. This last result needs to be supported by further studies. Simulations ``Z'', that use $\alpha=0.01$ -- i.e. doubled with respect to the ``A'' reference case -- do not appear to show significant differences from our reference case.

\section{Non-Axisymmetric Structures in Protostellar discs}\label{sec:nonaxfeat}

Recent high spatial resolution observations from near-infrared to mm and cm wavelengths have revealed spiral arms \citep{garufi2017, dong2018c}, cavities and gaps in the gas \citep{vandermarel2016, huang2018}, and cavities and gaps in the dust \citep{vandermarel2016, francis2020}.

A small selection of the discs with cavities, display large crescent shaped dust asymmetries \citep[e.g. ][]{vandermarel2013, casassus2015, tang2017}.
Despite the gas distribution being directly observable using molecular lines, highest resolution images from the SPHERE instrument on the VLT, ALMA and other observational facilities, are sensitive to the dust thermal emission, which may differ from the gas distribution. In particular, low contrast over-densities in the gas density structure act as pressure traps for dust grains (provided that they co-move with the flow), leading to perturbations in the dust density structure larger than those in the gas.

Three possible formation mechanisms can be invoked for these structures.

The first mechanism is a vortex co-moving with the gas (referred to as the ``Vortex scenario''). For this to operate the Rossby-Wave instability (RWI) must be triggered \citep{lovelace99,li2000,lovelace2014}. The RWI manifests whenever a strong gradient in the vortensity profile is present, provided the disc viscosity is sufficiently low ($\alpha_{\rm ss}\lesssim 10^{-4}$). This promotes the whirling motion of adjacent shearing layers, similar to the Kelvin-Helmholtz instability.
So-called ``dead zones'' in the disc have also been shown to cause density gradients affecting the vortensity in such a way that RWI is triggered \citep{regaly2012,ruge2016}. Numerical simulations have shown that the presence of a planet affects the vortensity gradient, causing some gas to accumulate outside its orbit, and carving a steep gap; for sufficiently low viscosities, the vortensity gradient is steep enough to enable the formation of a vortex.

A second formation mechanism for these dust asymmetries is the orbiting over-dense lump discussed earlier in this work (Sec. \ref{sec:horseform}). It involves the presence of a (sub-)stellar companion orbiting the primary star (secondary to primary mass ratio $M_{\rm 2}/M_{\rm 1}\gtrsim 0.2$), producing a poorly understood instability: the cavity size grows significantly and an over-dense feature orbiting at the edge of the cavity with Keplerian motion (i.e. co-moving with the gas) forms. No whirling motion is observed in this scenario \citep{ragusa2017,calcino2019}.

Both the over-dense lumps and vortices are expected to trap dust particles with growing efficiency when the gas-dust coupling is marginal (\citealp{birnstiel13a}, \citealp{vandermarel2021}) -- i.e. when the particle Stokes number approaches ${\rm St\sim 1}$. Indeed, both features are pressure maxima co-moving with the flow, which trap dust. However, in the vortex scenario, the growing dust-to-gas ratio within the the pressure maximum is expected to destroy the vortex \citep{johansen04a,fu2014}, even though 3D simulations fail to reproduce this effect \citep{lyra2018}.

A third mechanism to explain the origin of dust asymmetries has been discussed by \citet{ataiee2013}, the ``Traffic jam'' scenario (also discussed in this paper, Sec. \ref{sec:horseform}). In this scenario the presence of a planet increases the eccentricity of gas orbits; the clustering of eccentric orbits and the slowdown that the gas experiences when approaching their apocentre, causes the formation of an over-density which is slowly precessing (not orbiting) at the same rate as the pericentre longitude of the gas orbits. This mechanism does not produce over-densities with high enough contrast-ratios to be responsible for the formation of azimuthal over-dense features in discs -- our simulations show that this is the case even with cavities as eccentric as those produced by (sub)stellar companions.

Features produced by this mechanism are long lasting, as long as the disc maintain an eccentricity gradient. But they move with the frequency of the cavity longitude of pericentre -- in contrast to the two scenarios we previously described, where the feature spans the cavity edge with Keplerian frequency. Finally, eccentric features are not expected to trap dust grains, as they are actually ``traffic jams'' due to the the dust particles streamlines clustering at the apocentre of the orbit rather than particles being trapped in the over-density. As a consequence, the reasoning applied to the two previous scenarios where low amplitude gas perturbations could produce high contrast dust over-densities cannot be applied in this case.

\subsection{Observational implications}\label{sec:observ}

The most immediate observational consequence this work suggests is that co-planar discs surrounding binaries with sufficiently high mass ratios ($q\geq 0.05$) are expected to be significantly eccentric. As soon as the disc reaches the 1:2 resonance ($R=1.59\,a_{\rm bin}$) the disc eccentricity grows rapidly and the cavity size grows. This pushes material outwards producing an orbiting over-dense lump. In real discs, this condition is met in two cases; after the cloud core collapse, when the newly formed binary starts depleting the cavity area, and when secondary accretion events take place.

Our simulations suggest that a high contrast orbiting over-dense lump (Keplerian orbit, contrast ratio $\delta_\phi\gtrsim 10$) will last for a limited number of orbits ($\sim 1000$ orbits), leaving in its place an eccentric ``traffic jam'' (moving at the cavity pericentre precession rate, contrast ratio $\delta_\phi\approx 3\textrm{--}4$). If the cavity is eccentric, the traffic jam feature is always present, possibly leading to the formation of two distinct azimuthal structures: one orbiting at the cavity edge, the other fixed at the cavity apocentre.

Fig. \ref{fig:snapshotsHorse} and \ref{fig:periodogram} show that a low contrast gas over-density ($\delta_\phi\sim 1.5\textrm{--}2$) is expected to survive at longer timescales -- this is consistent with previous results in the literature where low contrast over-densities have been observed orbiting at the cavity edge after $t\sim 6000 \,t_{\rm orb}$ \citep{miranda2017}. Over-dense lumps co-moving with the flow (i.e. orbiting at the edge of the cavity) are effective in trapping dust grains starting from very small azimuthal contrast ratios ($\delta_\phi\gtrsim 1$, \citealp{birnstiel13a}; Van der Marel et al., in prep); this implies that high contrasts in the dust distribution, as those observed by ALMA, can be achieved starting from low contrast over-densities in the gas -- such as those weak orbiting features that survive at late timescales in our simulations.

There are a few ways to distinguish orbiting lumps and traffic jams observationally. First, as discussed above, since eccentric traffic jams cannot trap dust grains, the dust contrast ratio in these features is not expected to exceed the contrast in the gas. This implies that detecting azimuthal over-densities in the dust thermal emission with $\delta_\phi> 3\textrm{--}4$ excludes traffic jams as a possible scenario originating the feature. Second, the co-moving over-dense lump is expected to trap dust, therefore detecting dust trapping (different contrasts $\delta_\phi$ at different wavebands) can also help distinguishing the scenarios. Finally, repeated observations can be also used. Since the over-density explored in this work is co-moving, its orbital motion can be detected (e.g. \citealp{tuthill2002}). The traffic jam scenario produces a fixed feature which moves on a much longer time-scale (hundreds of binary orbits).

Despite high contrast over-densities not lasting for times longer than $t\gtrsim 1000 \,t_{\rm orb}$, for binaries opening cavities as large $a_{\rm cav}\gtrsim 100$ au, thousands of binary orbits correspond to time-scales $t\gtrsim 10^5$ yr, which is $\gtrsim10$\% of typical protoplanetary disc lifetimes \citep{haisch2001,kraus2012b,harris2012}. Since these features have been observed to form when the cavity is carved, we expect young systems to be more likely to show gas over-densities with $\delta_\phi\gtrsim 4$.

Secondary accretion events, flybys, or other later perturbations of the systems, could still in principle explain the presence of high contrast orbiting over-dense lumps also in the gas component of older systems.

The initial conditions in real physical systems are far from a steady state \citep{bate2018}. It is thus important to understand the evolution of the disc, not just when it has reached the quasi-steady state.

Vortices differ from orbiting over-dense lumps for the fact they might form also in systems where no cavity is present. However, they are expected to share most of the characterising features of orbiting over-dense lumps -- they also co-move with the flow and trap dust grains, causing high contrast dust over-densities. When a cavity is present, being able to detect the whirling motion of vortices through kinematic maps appear to be the only strategy to distinguish them from orbiting over-dense lumps. Nevertheless, no molecular lines observations with the required spatial resolution for this purpose are available yet.

\section{Summary and  conclusion}\label{sec:conclusion}

We performed a suite of of 3D SPH simulations of binaries surrounded by circumbinary discs. Our results suggest that most circumbinary discs with sufficiently high mass ratios, $q\geq 0.05$, develop an eccentric cavity, consistently with previous results in the literature \citep{farris2014,dorazio2016,miranda2017,munoz2020b}. The formation of eccentric cavities occurs over a wide range of disc+binary initial conditions, even though we always start with a circular binary and circular disc.

Our results suggest that:
\begin{enumerate}
\item The growth of the disc eccentricity appears to be driven by an unstable positive feedback mechanism involving the eccentric Lindblad resonance $m=3,\, l=2$, or circular $m=1,\, l=1$ -- that were previously suggested to be responsible for the growth of the disc and binary eccentricity \citep{dangelo2006,macfadyen2008}. However, the action of a non-resonant mechanism for disc eccentricity growth cannot be fully excluded.


\item Despite resonances being able to explain the evolution of the disc eccentricity, resonant binary-disc interaction theory alone, as we know it, seems not to be sufficient to explain the formation of cavities as large as $a_{\rm cav}=3.5\, a_{\rm bin}$ when the binary eccentricity is as low as $e\lesssim 0.01$. Resonances at those radii are not strong enough to overcome the viscous diffusion of the disc. The binary eccentricity required for that purpose is much higher than the one binaries develop in this work (see Fig. \ref{fig:resostrength}).
We speculate that ELRs increase their intrinsic strength with increasing disc eccentricity (in addition to the already well known and discussed dependence on the binary eccentricity). We also speculate that a non-resonant mechanism might be active carving the cavity (see also point \ref{point4} and \ref{point5} in this section, and Sec. \ref{sec:resostrength} for more details).

\item The maximum value of disc eccentricity that our simulations reach is set by the orbit crossing limit (see Fig. \ref{fig:ecrit}).
This limit is not related to the maximum value of eccentricity, but rather the steepness of the eccentricity profile. If the eccentricity gradient ($de/da$) is too large, the eccentric orbits of fluid elements intersect, resulting in shocks that prevent further growth of the eccentricity. In our simulations, discs around binaries with mass ratios $q>0.2$ undergo orbit intersection, reaching the same maximum value of disc eccentricity. For lower binary mass ratios, viscosity acts to damp the disc eccentricity.
When this effect balances the pumping action of resonances, the eccentricity stops growing. Thus, for binary mass ratios $q\leq 0.2$, since the intrinsic strength of resonances grows with the mass ratio, for a fixed value of disc viscosity, the higher the binary mass ratio is, the higher the maximum eccentricity. This could potentially lead to constraints on the disc viscosity for systems where the binary mass ratio and disc scale height can be measured.

\item \label{point4} Our analysis of disc eccentricity and cavity semi-major axis (cavity size) evidenced that these two quantities show an interesting linear correlation, which appear to be the same for all the simulations we examined (Fig. \ref{fig:correacav}). We believe this result is very important, as it constitutes a starting point for future developments of this work.

\item \label{point5} The pericentre radius of the cavity remains approximately constant throughout the entire length of the simulation, suggesting that the tidal torque sets the pericentre radius of the cavity non-resonantly and, as a consequence, the cavity semi-major axis grows due to the growth of the disc eccentricity.
In the light of point iv), this points in the direction of a non-resonant truncation mechanism being responsible for carving the cavity.
\end{enumerate}

Our simulations confirm some evolutionary features previously observed:
\begin{enumerate}

\item When the disc becomes eccentric, the material flows on nested elliptic orbits with decreasing eccentricity profile and aligned pericentres. The disc precesses rigidly, meaning that the elliptical orbits all precess together at the same rate, conserving the alignment of the pericentres \citep{macfadyen2008,teyssandier2016,miranda2017,ragusa2018,munoz2020b}. Eccentric cavities all show a ``traffic jam'' over-dense feature due to the clustering of nested eccentric orbits at the apocentre.

\item Simulations with mass ratio $q> 0.2$ show the formation of an azimuthal over-dense feature with $\delta_\phi\gtrsim 10$ -- known in the black hole binary community as ``over-dense lump'' and as ``horseshoe'' feature in the protoplanetary one \citep{shi2012,farris2014,ragusa2017,miranda2017} -- that orbits with Keplerian frequency at the edge of the cavity, produced by the strong tidal streams \citep{shi2012,dorazio2013} thrown by the binary.

Our results add to this picture that the fast growth of the cavity size appears to be one of the key ingredient for the formation of an high ($\delta_\phi\gtrsim 10$) contrast ratio over-dense co-moving feature. As soon as the cavity stops growing, the over-density also stops growing.

\item  As soon as a quasi-steady state configuration is reached (after $\approx 1000$ binary orbits) the disc progressively evolves towards a configuration with a slowly precessing ``eccentric traffic jam'' feature at the apocentre of the cavity $\delta_\phi\approx 3\textrm{--}4$ and a lower contrast orbiting over-dense lump ($\delta_\phi\approx1.5\textrm{--}2$) that co-moves with the flow. This result is consistent with what previously found by \citet{miranda2017}, who, for circular binaries, found an orbiting over-dense lump with $\delta_\phi\sim \textrm{2--3}$ is still present after 6000 binary orbits.
\end{enumerate}

We draw the following conclusions for observations of protoplanetary discs. Both high contrast and low contrast gas structures can lead to the formation of high contrast ratio features in the dust density field provided they co-move with the flow \citep{birnstiel13a}. This implies that our simulations with $q>0.2$ are all in principle consistent with hosting high contrast dust density structures for at least 2000 binary orbits, if dust was included in our simulations.
For typical systems, such timescale represents a significant fraction of their lifetime, we discuss this in Sec. \ref{sec:observ}. The contrast of this feature depends more on how much material is pushed outward when the cavity becomes eccentric rather than the value of the binary mass ratio, so that $q> 0.2$ should not be considered as a threshold for an high contrast orbiting over-dense lumps to form.

Reliable initial conditions of real physical systems are still very poorly constrained. Given the relatively long time-scales involved, the chances of observing a system while it is still relaxing towards a steady configuration are high, particularly in young protostellar systems.


High resolution kinematic data in protoplanetary discs can be used to test our theoretical results.

Future theoretical developments of this project involve a better investigation of the strength of ELRs in eccentric discs, and a strategy to understand the evolution of discs surrounding high mass ratio binaries on longer time-scales.

\section*{Acknowledgements}

We thank the anonymous referee for his/her insightful comments that substantially improved the conclusions of the paper.
ER thanks Jean Teyssandier, Giuseppe Lodato, Andrew King, Sergei Nayakshin, Diego Mu\~noz and Nienke van der Marel for fruitful discussion.
ER and RA acknowledge financial support from the European Research Council (ERC) under the European Union's Horizon 2020 research and innovation programme (grant agreement No 681601).
JC acknowledges support from an Australian Government Research Training Program.
This project has received funding from the European Union’s Horizon 2020 research and innovation programme under the Marie Skłodowska-Curie grant agreement No 823823 (DUSTBUSTERS).
The simulations performed for this paper ran for a total of $\sim 10^6$ cpu hours using the DiRAC Data Intensive service at Leicester, operated by the University of Leicester IT Services, which forms part of the STFC DiRAC HPC Facility (www.dirac.ac.uk). The equipment was funded by BEIS capital funding via STFC capital grants ST/K000373/1 and ST/R002363/1 and STFC DiRAC Operations grant ST/R001014/1. DiRAC is part of the National e-Infrastructure.
Fig. \ref{fig:simAs}, \ref{fig:snapshotsHorse} and \ref{fig:snapshots} were created using \textsc{splash} \citep{price07a}. All the other figures were created using \textsc{matplotlib} python library \citep{hunter2007}.

\section*{Data Availability Statement}

The \textsc{phantom} SPH code is available at \url{https://github.com/danieljprice/phantom}. The input files for generating our SPH simulations and analysis routines are available upon request.



\bibliographystyle{mnras}
\bibliography{biblio}




\label{lastpage}
\end{document}